\documentclass[12pt,a4paper]{article}
\usepackage{amsfonts}
\usepackage{amssymb}
\usepackage{amsmath}
\usepackage{graphicx}
\usepackage{subfigure}
\usepackage{multirow}
\usepackage[figuresright]{rotating}
\renewcommand{\today}{\number\year . \number\month .  \number\day .  }
\usepackage{booktabs}
\usepackage{subcaption}
\usepackage{url}
\usepackage{lineno}
\usepackage{url}
\usepackage{paralist}
\usepackage{xcolor}
\usepackage{listings}
\usepackage[chapter, nottoc]{tocbibind} 
\usepackage{wrapfig}
\usepackage{verbatim}
\usepackage[figuresright]{rotating}

\usepackage{lscape}
\usepackage{makecell}
\usepackage{booktabs}
\usepackage{tablefootnote}
\usepackage{subcaption}

\usepackage{ifthen} % for conditional statements
\newboolean{pdflatex}
\setboolean{pdflatex}{true} % False for eps figures 

\newboolean{articletitles}
\setboolean{articletitles}{true} % False removes titles in references

\newboolean{uprightparticles}
\setboolean{uprightparticles}{false} %True for upright particle symbols
\usepackage[en-US]{datetime2}
\def\paperauthors{LHCb collaboration} % Leave as is for PAPER, CONF and FIGURE
\def\paperasciititle{Study of Lambda_b and Xi_b decays to Lambda0 h+ h- and evidence for CP violation in Lambda_b to Lambda0 K+ K- decays} % Set ASCII title here !! MAKE sure it's only ASCII characters !! 
\def\papertitle{Study of  $\Lb$ and $\Xibz$ decays to $\Lz h^{+} h^{\prime -} $ and evidence for \CP violation in $\Lb\to\Lz \Kp\Km$ decays} % Latex formatted title
\def\paperkeywords{{High Energy Physics}, {LHCb}} % Comma separated list
\def\papercopyright{\the\year\ CERN for the benefit of the LHCb collaboration} % new since 9/Apr/2018
\def\paperlicence{CC BY 4.0 licence}
\def\paperlicenceurl{https://creativecommons.org/licenses/by/4.0/}

\newif\ifEnableSectionTOCLinks
\EnableSectionTOCLinksfalse % deactivated
\usepackage[top=1in, bottom=1.25in, left=1in, right=1in]{geometry}

\usepackage{microtype}
\usepackage{lineno}  % for line numbering during review
\usepackage{xspace} % To avoid problems with missing or double spaces after
\usepackage{caption} %these three command get the figure and table captions automatically small

\usepackage{graphicx}  % to include figures (can also use other packages)
\usepackage{color}
\usepackage{colortbl}
\graphicspath{{./figs/}} % Make Latex search fig subdir for figures
\usepackage{amsmath} % Adds a large collection of math symbols
\usepackage{amssymb}
\usepackage{amsfonts}
\usepackage{upgreek} % Adds in support for greek letters in roman typeset

\newcommand*\patchAmsMathEnvironmentForLineno[1]{%
\expandafter\let\csname old#1\expandafter\endcsname\csname #1\endcsname
\expandafter\let\csname oldend#1\expandafter\endcsname\csname
end#1\endcsname
 \renewenvironment{#1}%
   {\linenomath\csname old#1\endcsname}%
   {\csname oldend#1\endcsname\endlinenomath}%
}
\newcommand*\patchBothAmsMathEnvironmentsForLineno[1]{%
  \patchAmsMathEnvironmentForLineno{#1}%
  \patchAmsMathEnvironmentForLineno{#1*}%
}
\AtBeginDocument{%
\patchBothAmsMathEnvironmentsForLineno{equation}%
\patchBothAmsMathEnvironmentsForLineno{align}%
\patchBothAmsMathEnvironmentsForLineno{flalign}%
\patchBothAmsMathEnvironmentsForLineno{alignat}%
\patchBothAmsMathEnvironmentsForLineno{gather}%
\patchBothAmsMathEnvironmentsForLineno{multline}%
\patchBothAmsMathEnvironmentsForLineno{eqnarray}%
}

\usepackage{hyperxmp}

\usepackage[pdftex,
            pdfauthor={\paperauthors},
            pdftitle={\paperasciititle},
            pdfkeywords={\paperkeywords},
            pdfcopyright={Copyright (C) \papercopyright},
            pdflicenseurl={\paperlicenceurl}]{hyperref}
\usepackage[colorinlistoftodos,textsize=scriptsize]{todonotes}

\usepackage[bottom,flushmargin,hang,multiple]{footmisc}

\usepackage[all]{hypcap} % Internal hyperlinks to floats.

\usepackage{xspace} 
\usepackage{upgreek}

\def\lhcb   {\mbox{LHCb}\xspace}

\def\besiii {\mbox{BESIII}\xspace}

\def\lhc    {\mbox{LHC}\xspace}

\def\rich   {RICH\xspace}

\def\MagUp {\mbox{\em Mag\kern -0.05em Up}\xspace}

\ifthenelse{\boolean{uprightparticles}}%
{

 \def\Pmu         {\ensuremath{\upmu}\xspace}                 
 \def\Pnu         {\ensuremath{\upnu}\xspace}                 
                  
 \def\Ppi         {\ensuremath{\uppi}\xspace}

 \def\Ppsi        {\ensuremath{\uppsi}\xspace}

 \def\PDelta      {\ensuremath{\Delta}\xspace}                 
 \def\PXi         {\ensuremath{\Xi}\xspace}                 
 \def\PLambda     {\ensuremath{\Lambda}\xspace}                 
 \def\PSigma      {\ensuremath{\Sigma}\xspace}                 
 \def\POmega      {\ensuremath{\Omega}\xspace}                 
 \def\PUpsilon    {\ensuremath{\Upsilon}\xspace}
 \let\oldPi\Pi
 \def\PPi         {\ensuremath{\oldPi}\xspace}

 \def\PB      {\ensuremath{\mathrm{B}}\xspace}                 
 \def\PD      {\ensuremath{\mathrm{D}}\xspace}                 
 \def\PJ      {\ensuremath{\mathrm{J}}\xspace}                 
 \def\PK      {\ensuremath{\mathrm{K}}\xspace}                 
 \def\Pb      {\ensuremath{\mathrm{b}}\xspace}                 
 \def\Pc      {\ensuremath{\mathrm{c}}\xspace}

 \def\Pp      {\ensuremath{\mathrm{p}}\xspace}                 

 \def\Ps      {\ensuremath{\mathrm{s}}\xspace}

 \def\thebaroffset{0.0em}
}
{

 \def\Pmu         {\ensuremath{\mu}\xspace}                 
 \def\Pnu         {\ensuremath{\nu}\xspace}                 
                  
 \def\Ppi         {\ensuremath{\pi}\xspace}

 \def\Ppsi        {\ensuremath{\psi}\xspace}                 
                  
 \mathchardef\PDelta="7101
 \mathchardef\PXi="7104
 \mathchardef\PLambda="7103
 \mathchardef\PSigma="7106
 \mathchardef\POmega="710A
 \mathchardef\PUpsilon="7107
 \mathchardef\PPi="7105
 \def\PB      {\ensuremath{B}\xspace}                 
 \def\PD      {\ensuremath{D}\xspace}                 
 \def\PJ      {\ensuremath{J}\xspace}                 
 \def\PK      {\ensuremath{K}\xspace}                 
 \def\Pb      {\ensuremath{b}\xspace}                 
 \def\Pc      {\ensuremath{c}\xspace}

 \def\Pp      {\ensuremath{p}\xspace}                 

 \def\Ps      {\ensuremath{s}\xspace}

 \def\thebaroffset{0.18em}
}
\newcommand{\offsetoverline}[2][\thebaroffset]{\kern #1\overline{\kern -#1 #2}}%

\makeatletter
\ifcase \@ptsize \relax% 10pt
  \newcommand{\miniscule}{\@setfontsize\miniscule{4}{5}}% \tiny: 5/6
\or% 11pt
  \newcommand{\miniscule}{\@setfontsize\miniscule{5}{6}}% \tiny: 6/7
\or% 12pt
  \newcommand{\miniscule}{\@setfontsize\miniscule{5}{6}}% \tiny: 6/7
\fi
\makeatother

\DeclareRobustCommand{\optbar}[1]{\shortstack{{\miniscule (\rule[.5ex]{1.25em}{.18mm})}
  \\ [-.7ex] $#1$}}

\def\mup        {{\ensuremath{\Pmu^+}}\xspace}
\def\mun        {{\ensuremath{\Pmu^-}}\xspace} % muon negative (\mum is taken)

\def\mumu       {{\ensuremath{\Pmu^+\Pmu^-}}\xspace}

\def\neub       {{\ensuremath{\overline{\Pnu}}}\xspace}

\def\neumb      {{\ensuremath{\neub_\mu}}\xspace}
\def\squark    {{\ensuremath{\Ps}}\xspace}

\def\cquark    {{\ensuremath{\Pc}}\xspace}

\def\bquark    {{\ensuremath{\Pb}}\xspace}

\def\pion   {{\ensuremath{\Ppi}}\xspace}

\def\pip    {{\ensuremath{\pion^+}}\xspace}
\def\pim    {{\ensuremath{\pion^-}}\xspace}
\def\pipm   {{\ensuremath{\pion^\pm}}\xspace}

\def\kaon    {{\ensuremath{\PK}}\xspace}
\def\KorKbar {\kern \thebaroffset\optbar{\kern -\thebaroffset \PK}{}\xspace}

\def\Kp      {{\ensuremath{\kaon^+}}\xspace}
\def\Km      {{\ensuremath{\kaon^-}}\xspace}

\def\KS      {{\ensuremath{\kaon^0_{\mathrm{S}}}}\xspace}

\def\D       {{\ensuremath{\PD}}\xspace}

\def\DorDbar {\kern \thebaroffset\optbar{\kern -\thebaroffset \PD}\xspace}
\def\Dz      {{\ensuremath{\D^0}}\xspace}

\def\Dp      {{\ensuremath{\D^+}}\xspace}
\def\Dm      {{\ensuremath{\D^-}}\xspace}

\def\DpDm    {\ensuremath{\Dp {\kern -0.16em \Dm}}\xspace}

\def\Dstarp  {{\ensuremath{\D^{*+}}}\xspace}

\def\B       {{\ensuremath{\PB}}\xspace}

\def\BorBbar {\kern \thebaroffset\optbar{\kern -\thebaroffset \PB}\xspace}

\def\Bd      {{\ensuremath{\B^0}}\xspace}

\def\BdorBdbar {\kern \thebaroffset\optbar{\kern -\thebaroffset \Bd}\xspace}

\def\Bpm     {{\ensuremath{\B^\pm}}\xspace}

\def\Bs      {{\ensuremath{\B^0_\squark}}\xspace}

\def\BsorBsbar {\kern \thebaroffset\optbar{\kern -\thebaroffset \Bs}\xspace}

\def\jpsi     {{\ensuremath{{\PJ\mskip -3mu/\mskip -2mu\Ppsi}}}\xspace}

\def\Y#1S{\ensuremath{\PUpsilon{(#1S)}}\xspace}

\def\proton      {{\ensuremath{\Pp}}\xspace}
\def\antiproton  {{\ensuremath{\overline \proton}}\xspace}

\def\Lz          {{\ensuremath{\PLambda}}\xspace}
\def\Lbar        {{\ensuremath{\offsetoverline{\PLambda}}}\xspace}
\def\LorLbar     {\kern \thebaroffset\optbar{\kern -\thebaroffset \PLambda}\xspace}

\def\Xires       {{\ensuremath{\PXi}}\xspace}

\def\Xiresbar       {{\ensuremath{\offsetoverline{\Xires}}}\xspace}

\def\Lc          {{\ensuremath{\Lz^+_\cquark}}\xspace}
\def\Lcbar       {{\ensuremath{\Lbar{}^-_\cquark}}\xspace}

\def\Xicp        {{\ensuremath{\Xires^+_\cquark}}\xspace}

\def\Lb           {{\ensuremath{\Lz^0_\bquark}}\xspace}
\def\Lbbar        {{\ensuremath{\Lbar{}^0_\bquark}}\xspace}

\def\Xibz         {{\ensuremath{\Xires^0_\bquark}}\xspace}
\def\Xibm         {{\ensuremath{\Xires^-_\bquark}}\xspace}

\def\Xibbarz      {{\ensuremath{\Xiresbar{}_\bquark^0}}\xspace}

\def\BF         {{\ensuremath{\mathcal{B}}}\xspace}

\newcommand{\decay}[2]{\ensuremath{#1\!\to #2}\xspace} 

\def\to                 {\ensuremath{\rightarrow}\xspace}

\def\CP                {{\ensuremath{C\!P}}\xspace}

\newcommand{\ACP}{{\ensuremath{{\mathcal{A}}^{\CP}}}\xspace}
\newcommand{\ACPraw}{{\ensuremath{{\mathcal{A}}^{\CP}}_{\text{raw}}}\xspace}

\def\AT#1     {\ensuremath{A_{\mathrm{T}}^{#1}}\xspace}           % 2

\def\C#1      {\ensuremath{\mathcal{C}_{#1}}\xspace}                       % 9
\def\Cp#1     {\ensuremath{\mathcal{C}_{#1}^{'}}\xspace}                    % 7
\def\Ceff#1   {\ensuremath{\mathcal{C}_{#1}^{\mathrm{(eff)}}}\xspace}        % 9  
\def\Cpeff#1  {\ensuremath{\mathcal{C}_{#1}^{'\mathrm{(eff)}}}\xspace}       % 7
\def\Ope#1    {\ensuremath{\mathcal{O}_{#1}}\xspace}                       % 2
\def\Opep#1   {\ensuremath{\mathcal{O}_{#1}^{'}}\xspace}                    % 7

\newcommand{\aunit}[1]{\ensuremath{\text{\,#1}}}       
\newcommand{\tev}{\aunit{Te\kern -0.1em V}\xspace}
\newcommand{\gev}{\aunit{Ge\kern -0.1em V}\xspace}
\newcommand{\mev}{\aunit{Me\kern -0.1em V}\xspace}
\newcommand{\kev}{\aunit{ke\kern -0.1em V}\xspace}
\newcommand{\ev}{\aunit{e\kern -0.1em V}\xspace}
 
\newcommand{\mevc}{\ensuremath{\aunit{Me\kern -0.1em V\!/}c}\xspace}
\newcommand{\gevc}{\ensuremath{\aunit{Ge\kern -0.1em V\!/}c}\xspace}
\newcommand{\mevcc}{\ensuremath{\aunit{Me\kern -0.1em V\!/}c^2}\xspace}
\newcommand{\gevcc}{\ensuremath{\aunit{Ge\kern -0.1em V\!/}c^2}\xspace}
\def\fb   {\ensuremath{\aunit{fb}}\xspace}
\def\invfb   {\ensuremath{\fb^{-1}}\xspace}

\def\gsim{{~\raise.15em\hbox{$>$}\kern-.85em
          \lower.35em\hbox{$\sim$}~}\xspace}
\def\lsim{{~\raise.15em\hbox{$<$}\kern-.85em
          \lower.35em\hbox{$\sim$}~}\xspace}

\def\sPlot{\mbox{\em sPlot}\xspace}

\def\sqs   {\ensuremath{\protect\sqrt{s}}\xspace}

\def\pt         {\ensuremath{p_{\mathrm{T}}}\xspace}

\def\evtgen     {\mbox{\textsc{EvtGen}}\xspace}

\def\geant      {\mbox{\textsc{Geant4}}\xspace}

\def\photos     {\mbox{\textsc{Photos}}\xspace}

\def\pythia     {\mbox{\textsc{Pythia}}\xspace}

\def\tell1  {TELL1\xspace}
\def\ukl1   {UKL1\xspace}

\newcommand{\lhcborcid}[1]{\href{https://orcid.org/#1}{\hspace*{0.1em}\raisebox{-0.45ex}{\includegraphics[width=1em]{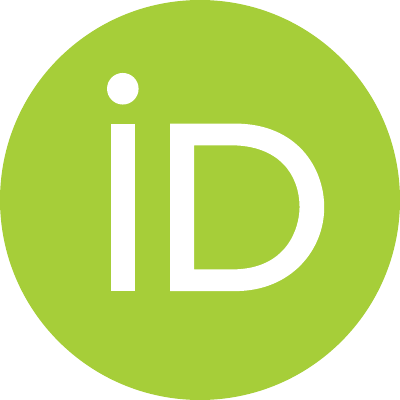}}}}

\def\LzKK       {{$\decay{\Lb}{\Lz K^+K^-}$}\xspace}
\def\LzKP       {{$\decay{\Lb}{\Lz K^+\pi^-}$}\xspace}
\def\LzPP       {{$\decay{\Lb}{\Lz \pi^+\pi^-}$}\xspace}

\def\XiPP       {{$\decay{\Xibz}{\Lz \pi^+\pi^-}$}\xspace}
\def\XiPK       {{$\decay{\Xibz}{\Lz \Km \pip}$}\xspace}
\def\XiKK       {{$\decay{\Xibz}{\Lz K^+K^-}$}\xspace}
\hypersetup{
  colorlinks   = true, %Colours links instead of ugly boxes
  urlcolor     = blue, %Colour for external hyperlinks
  linkcolor    = blue, %Colour of internal links
  citecolor    = red   %Colour of citations
}

\ifEnableSectionTOCLinks
    \usepackage[explicit]{titlesec} % to change headings
    
    \let\oldcontentsline\contentsline
    \renewcommand

    \titleformat{\section}{\normalfont\Large\bf}{\hyperlink{tocsection.\thesection}{{\thesection} \parbox[t]{\dimexpr\textwidth-1pc}{#1}}}{1pc}{}

    \titleformat{\subsection}{\normalfont\bf}{\hyperlink{tocsubsection.\thesubsection}{{\thesubsection} \parbox[t]{\dimexpr\textwidth-1pc}{#1}}}{1pc}{}

    \titleformat{name=\section,numberless}[display]{}{}{0pt}{\normalfont\Huge\bfseries #1}
\fi

\usepackage{cite} % Allows for ranges in citations
\usepackage{mciteplus}
\begin{document}

%%%%%%%%%%%%%%%%%%%%%%%%%
%%%%% Title     %%%%%%%%%
%%%%%%%%%%%%%%%%%%%%%%%%%
\renewcommand{\thefootnote}{\fnsymbol{footnote}}
\setcounter{footnote}{1}

%%%%%%%% CHOOSE TITLE PAGE--------
% ===============================================================================
% Purpose: LHCb-PAPER journal paper title page template
% Author: 
% Created on: 2010-09-25
% ===============================================================================

%%%%%%%%%%%%%%%%%%%%%%%%%
%%%%%  TITLE PAGE  %%%%%%
%%%%%%%%%%%%%%%%%%%%%%%%%
\begin{titlepage}
\pagenumbering{roman}

% Header ---------------------------------------------------
\vspace*{-1.5cm}
\centerline{\large EUROPEAN ORGANIZATION FOR NUCLEAR RESEARCH (CERN)}
\vspace*{1.5cm}
\noindent
\begin{tabular*}{\linewidth}{lc@{\extracolsep{\fill}}r@{\extracolsep{0pt}}}
\ifthenelse{\boolean{pdflatex}}% Logo format choice
{\vspace*{-1.5cm}\mbox{\!\!\!\includegraphics[width=.14\textwidth]{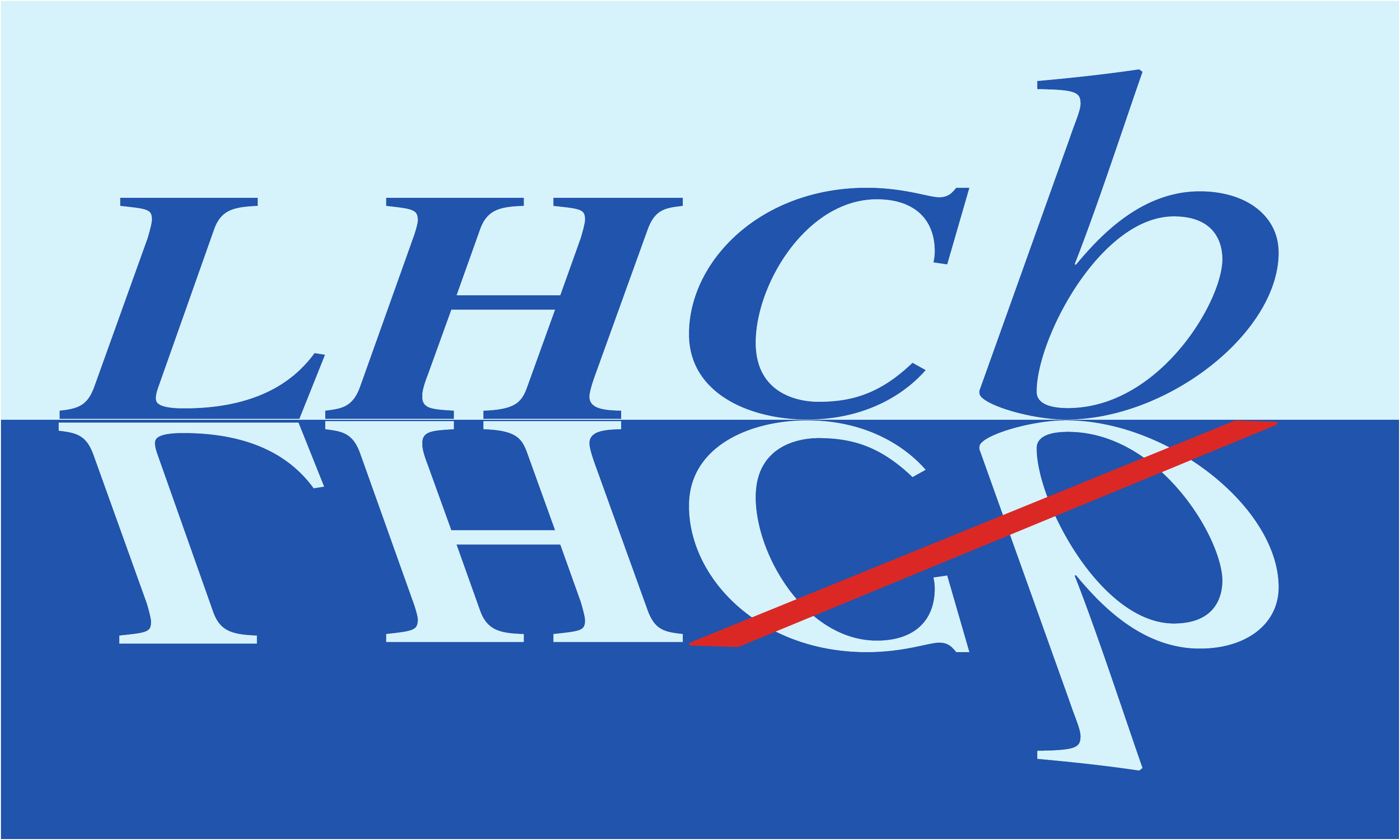}} & &}%
{\vspace*{-1.2cm}\mbox{\!\!\!\includegraphics[width=.12\textwidth]{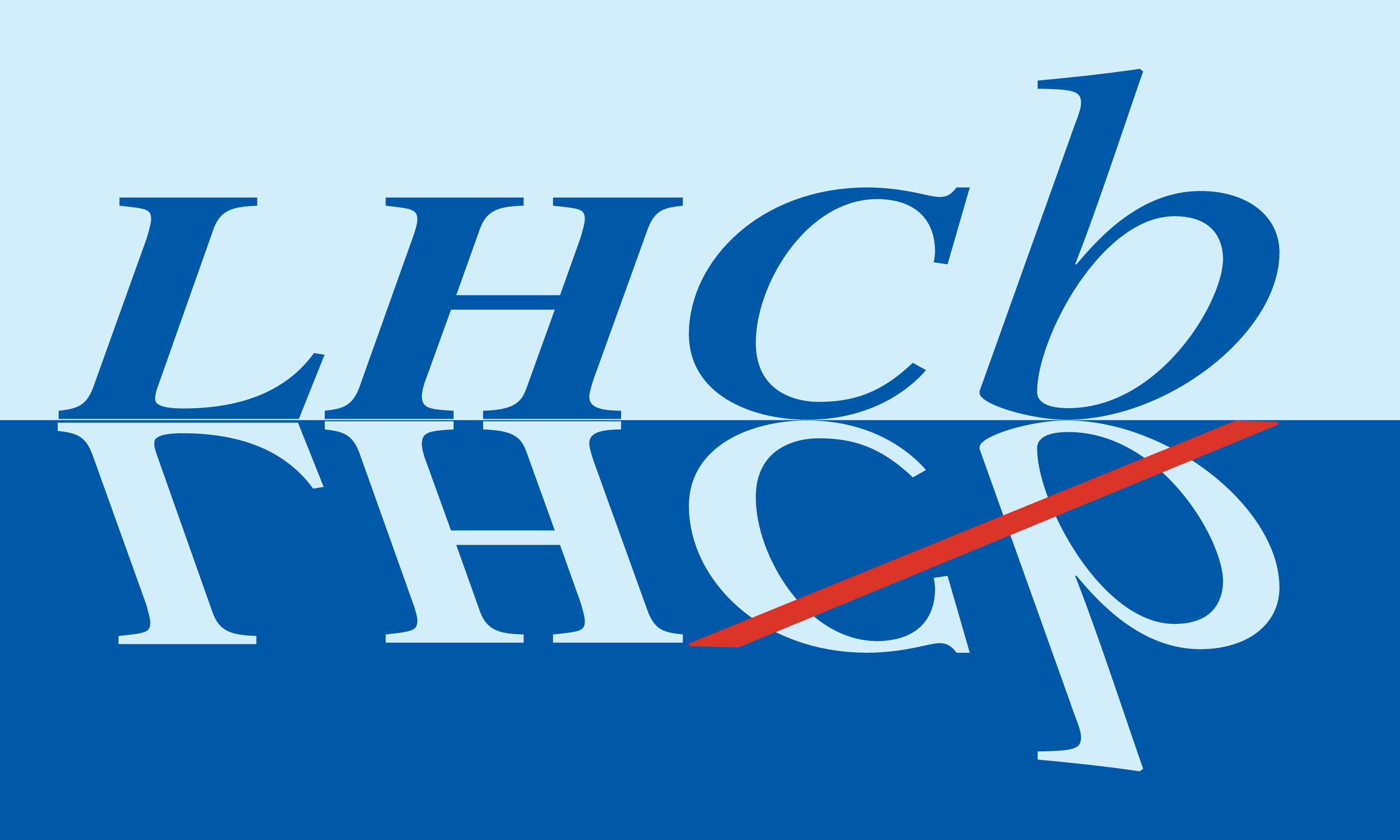}} & &}%
\\
 & & CERN-EP-2024-281 \\  % ID 
 & & LHCb-PAPER-2024-043 \\  % ID 
 & & \today \\ 
 & & \\
% not in paper \hline
\end{tabular*}

\vspace*{4.0cm}

{\normalfont\bfseries\boldmath\huge
\begin{center}
% DO NOT EDIT HERE. Instead edit macro in main.tex to keep metadata correct
  \papertitle 
\end{center}
}

\vspace*{2.0cm}

% Authors -------------------------------------------------
\begin{center}
%In the footnote, replace 'paper' by 'Letter' in case of submission to PRL or PLB 
\paperauthors\footnote{Authors are listed at the end of this paper.}
\end{center}

\vspace{\fill}

% Abstract -----------------------------------------------
\begin{abstract}
  \noindent
 A study of $\Lb$ and $\Xibz$ decays to $\Lz h^{+} h^{\prime -}$  $(h^{(\prime)}=\pi, K)$ is performed
using  $pp$ collision data collected by the LHCb experiment during LHC Runs 1--2, corresponding to an integrated luminosity of $9 \invfb$.
The  branching fractions for these decays are measured using the $\Lb\to\Lc(\to\Lz\pip)\pim$ decay as control channel. 
The decays \LzPP and \XiPK are observed for the first time.
For  decay modes with sufficient signal yields,  \CP\ asymmetries are measured in the full and localized regions of the final-state phase space. Evidence is found for \CP violation in the  $\Lb\rightarrow \Lz \Kp\Km$ decay, interpreted as originating primarily from an asymmetric $\Lb \to N^{*+} K^-$ decay amplitude. 
The measured \CP asymmetries for the other decays are compatible with zero. 
  
\end{abstract}

\vspace*{2.0cm}

\begin{center}
  Published in 
  Physical~Review~Letters~134~(2025)~101802
\end{center}

\vspace{\fill}

{\footnotesize 
\centerline{\copyright~\papercopyright. \href{\paperlicenceurl}{\paperlicence}.}}
\vspace*{2mm}

\end{titlepage}

%%%%%%%%%%%%%%%%%%%%%%%%%%%%%%%%
%%%%%  EOD OF TITLE PAGE  %%%%%%
%%%%%%%%%%%%%%%%%%%%%%%%%%%%%%%%

\newpage
\setcounter{page}{2}
\mbox{~}

\renewcommand{\thefootnote}{\arabic{footnote}}
\setcounter{footnote}{0}
\cleardoublepage

%%%%%%%%%%%%%%%%%%%%%%%%%
%%%%% Main text %%%%%%%%%
%%%%%%%%%%%%%%%%%%%%%%%%%

\pagestyle{plain} 
\setcounter{page}{1}
\pagenumbering{arabic}

% \linenumbers

%% This is the main body

In the Standard Model (SM) of particle physics, symmetry breaking under the combined charge-conjugation and parity transformations (\CP violation) originates from  a complex phase within the Cabibbo--Kobayashi--Maskawa (CKM) matrix~\cite{HFLAV21}. To date, all observed \CP violation phenomena 
align with the CKM mechanism.
However, the amount of \CP violation in the SM is insufficient to explain the observed matter-antimatter imbalance in the Universe~\cite{Riotto:1999yt}, motivating further study of \CP violation and  searches for possible new sources beyond the SM contributions. 

While the breaking of \CP symmetry has been established and extensively studied in $K$, $B$ and $D$ meson decays, it has never been observed in any baryon decay. The \besiii experiment has conducted comprehensive searches for \CP violation in light hyperon decays, including studies of decay rates and 
parameters, 
finding no evidence for \CP violation~\cite{BESIII:2022qax}.
Searches for \CP violation have been pursued by \lhcb 
in bottom-baryon decays, including $\decay{\Lb}{\KS p \pim}$~\cite{LHCb-PAPER-2013-061},\footnote{The inclusion of charge-conjugated processes is implied throughout the article if not specified.} $\decay{\Lb}{\jpsi p \pim}$~\cite{LHCb-PAPER-2014-020} $\decay{\Lb}{p h^{-}h^{\prime+}h^{\prime\prime-}}$\cite{LHCb-PAPER-2016-030,LHCb-PAPER-2018-001, LHCb-PAPER-2018-044, LHCb-PAPER-2019-028, LHCb-PAPER-2021-027}, $\mbox{\decay{\Lb}{\Lz \Kp\Km}}$, $\mbox{\decay{\Lb}{\Lz \Kp\pim}}$~\cite{LHCb-PAPER-2016-004}, $\decay{\Lb}{ p \Km \mup \mun}$~\cite{LHCb-PAPER-2016-059}, 
$\decay{\Lb}{p h^-}$~\cite{LHCb-PAPER-2018-025}, $\decay{\Xibm}{ p\Km\Km}$~\cite{LHCb-PAPER-2020-017},
\mbox{$\decay{\Lb}{\Lz\gamma}$}~\cite{LHCb-PAPER-2021-030}, $\decay{\Lb}{\Lz\phi}$~\cite{LHCb-PAPER-2016-002}, 
and charm-baryon decays such as $\decay{\Lc}{p h^+ h^-}$~\cite{LHCb-PAPER-2017-044} and 
 $\decay{\Xicp }{p \Km \pip}$~\cite{LHCb-PAPER-2019-026},  
 where 
 $h, h', h^{''}=\pi$ 
 or $K$ throughout this Letter. 
These measurements are statistically limited, as most of them use only 
data collected during \lhc Run~1 (2011--2012). 
Further investigation of \CP violation in baryon decays may shed new light on the dynamics of weak decays in the baryon sector and provide a better picture of \CP violation originating from quark transitions.

In three-body charmless  $\B$-meson decays,  $\decay{\B}{ h^+ h'^- h''^+}$, large \CP violation up to 75\% 
is observed in localized regions of phase space, for example in the low  $\Kp\Km$, low $\pip\pim$ and high $\pip\pim$ mass regions~\cite{LHCb-PAPER-2021-049, LHCb-PAPER-2018-051, LHCb-PAPER-2014-044}. 
These results suggest that resonance interactions and $\pip\pim \leftrightarrow \Kp\Km$ S-wave rescattering play an important role in the generation of  strong phases needed for direct \CP violation,
and motivate further studies of $\Lb$ and $\Xibz$ decays to $\Lz h^+h'^-$ final states, which are governed by similar  dynamics in the SM.

Quasi-two-body charmless $\Lb$ decays, $\decay{\Lb}{\Lz \omega/\Lz\phi/\Lz\rho}$, have been studied with the QCD factorization approach and their \CP violation is predicted to be in the range $0\%$ to $4\%$ with branching fractions at the level of $10^{-7}$~\cite{Geng:2016gul,PhysRevD.69.017901,Guo:1998eg,PhysRevD.99.054020}. The generalized factorization approach (GFA), considering part of the nonfactorizable sources by introducing an effective color number $N_c$, gives similar \CP asymmetry predictions but the branching fractions are predicted to be approximately $10^{-6}$~\cite{Rui:2022jff, PhysRevD.95.093001}. For the $\Lb\to N^{*+}\pim$ decay, the \CP asymmetry is predicted to be in the range from $-4\%$ to $6\%$~\cite{Wang:2024}. In a previous \lhcb study, the \LzKK and \LzKP decays were observed with the Run~1 sample~\cite{LHCb-PAPER-2016-004}, where the first evidence for \LzPP was established and the \CP asymmetries for these decays were found to be compatible with zero. 

Further higher-precision measurements of \CP asymmetries and branching fractions of $\Lb$ and $\Xibz$ decays to $\Lz h^+ h'^-$ final states offer stringent tests of these models and 
provide a foundation to study other quasi-two-body decays that have not been considered before.

This Letter reports the measurements of branching fractions and \CP violation parameters for charmless decays of $\Lb$ and $\Xibz$ baryons into the final states $\Lz K^\pm\pi^\mp$, $\Lz \Kp\Km$, and $\Lz \pip\pim$, among which the suppressed modes $\decay{\Lb}{\Lz \Km\pip}$ and $\decay{\Xibz}{\Lz \Kp\pim}$ are not considered. The study is performed based on proton-proton ($pp$) collision data collected with the \lhcb detector during \lhc 
Runs 1--2 (2011--2018) at center-of-mass energies of 7, 8 and 13\tev and corresponding to an integrated luminosity of 9$\invfb$.
The $\decay{\Lb}{\Lc(\to\Lz\pip) \pim}$ decay is used as control channel for both the branching fraction and \CP-violation measurements to reduce systematic uncertainties.

%%%%%%%%%%%% Data Acquisition %%%%%%%%%%%
The LHCb detector is a single-arm forward spectrometer covering the pseudorapidity range $2 < \eta < 5$, described in detail in Refs.~\cite{LHCb-DP-2008-001,LHCb-DP-2014-002}. It is designed specifically for the study of particles containing $b$ or $c$ quarks. Of particular relevance for this analysis is the tracking system, comprising silicon-strip stations upstream and straw drift tube stations downstream of a $4{\mathrm{\,T\,m}}$ dipole magnet~\cite{LHCb-DP-2013-003, LHCb-DP-2017-001}, and the Ring-Imaging Cherenkov (\rich)~\cite{LHCb-DP-2012-003} detectors used for the particle identication (PID)~\cite{LHCb-PUB-2016-021,LHCb-DP-2019-001}, whose performance of simulated samples is calibrated to match that evaluated with high-yield decay modes in data. 
The $\Lb/\Xibz$ decays are selected by an online trigger system which consists of a hardware stage followed by a software stage~\cite{LHCb-DP-2012-004,LHCb-DP-2019-001}. 
The hardware trigger is based on information from the calorimeter and muon systems. 
The software trigger applies full event reconstruction, selecting events with a two-, three- or four-track secondary vertex with a significant displacement from any
primary $pp$ interaction vertex. 
Simulated  $\decay{\Lb/\Xibz}{\Lz h^+h'^-}$ decays are used to model the effects of the detector acceptance and imposed selection requirements, and the signal mass distributions. 
In the simulation, samples are
generated with \pythia~\cite{Sjostrand:2007gs,*Sjostrand:2006za}, \evtgen~\cite{Lange:2001uf}, \photos~\cite{Golonka:2005pn} and the \geant toolkits~\cite{Allison:2006ve, *Agostinelli:2002hh} as described in Ref.~\cite{LHCb-PROC-2011-006}.

In the offline selection, tracks identified as a proton and a pion are used to form a $\Lz$ candidate, which is further combined with a pair of oppositely charged hadrons identified as a pion or kaon to form a $\Lb/\Xibz$ candidate.
Backgrounds from specific narrow resonances including $\KS$, $\Dz$, $\Lc$, $\Xicp$, $\jpsi$ and $\chi_{c0}$ hadrons formed by combinations of tracks from the final state particles of $\Lb/\Xibz$ candidates are removed by vetoing in the relevant mass spectra. 
Further discrimination of signal from background is achieved through a  Boosted Decision Tree (BDT) classifier~\cite{Breiman,Hocker:2007ht}, using a combination of  kinematic and topological variables as inputs.
The BDT classifier is trained with simulated $\decay{\Lb}{\Lz \pip \pim}$ decays as the signal, and using the data sample in the mass region $m(\Lz \pip \pim) \in [5800, 6100]\mevcc$  as the  background. 
Requirements on the BDT response and  PID of final-state tracks are optimized and applied simultaneously to maximize the figure-of-merit, defined as $N_{\rm{S}}/\sqrt{N_{\rm{S}} + N_{\rm{B}}}$ ($N_{\rm{S}}/(\sqrt{N_{\rm{B}}} + 2.5)$) for the $\Lb$ ($\Xibz$) decays. Here $N_{\rm{S}}$ and $N_{\rm{B}}$ are the signal and background yields in the $\Lb$ ($\Xibz$) signal region, defined as a $\pm 50\mevcc$ mass window around the known $\Lb$ ($\Xibz$) mass~\cite{PDG2024}. 
The PID requirements help to reduce combinatorial background and cross-feeds from other signal decays and from $B$-meson decays. The contributions from $B$-meson decays are suppressed to a negligible level. 
%%%%%%%%%%%%%%%%%%%%%% Fit %%%%%%%%%%%%%%%%%%%%%%%%

The $\Lz h^+ h'^-$ mass distributions after all selections are shown in Fig.~\ref{fig:mass spectrum} of the End Matter, together with fit projections. 
The obtained signal yields are summarized in Table~\ref{tab:branching_ratios}, extracted using a simultaneous unbinned maximum-likelihood fit to all the $\Lz h^+ h'^-$ mass distributions, where the two \CP conjugate states are combined. 
The signal component in the corresponding $\Lz h^+ h'^-$ mass distribution is modeled by the sum of two Crystal Ball (CB) functions~\cite{Skwarnicki:1986xj}, with tail parameters fixed from simulation. 
The distributions of cross-feeds from other signal decays due to misidentified $h^+$ or $h'^-$ hadrons are obtained from simulation, and their yields are constrained to the respective yields of the correctly reconstructed signals multiplied by the experimental efficiencies evaluated from simulation. 
The decay $\decay{\Lb}{\Lz h^+h'^-\gamma/\pi^0}$, with $\gamma/\pi^0$ not reconstructed, is modeled by an ARGUS function~\cite{ARGUS:1990hfq} convolved by a Gaussian distribution for the experimental resolution. The shape parameters of ARGUS function are constrained from simulation. 
The combinatorial background is modeled by an exponential function. For \CP-violation measurements, 
the signal and background parameters are shared between baryon and antibaryon decays.

Using Wilks' theorem~\cite{Wilks:1938dza}, the statistical significances of the \mbox{\LzPP} and \mbox{\XiPK} decays are measured to be more than $10\,\sigma$, giving the first observation of these decays. The
significance of the \XiPP decay is determined to be $4\,\sigma$, while that of the \XiKK decay is about $1.7\,\sigma$.

\begin{table}[!t]
\centering
\caption{Signal yield and  (upper limit of) \CP-averaged branching fraction (\BF)  for each decay mode. The uncertainties are statistical, systematic and due to the branching fraction of the control mode. The yield for the control mode is also shown.}
\label{tab:branching_ratios}
\begin{tabular}{@{}lcc@{}}
\hline
Decay &  Yield   & $\BF~(\times 10^{-6})$ \\ \hline
$\decay{\Lb}{\Lz\pip \pim}$ &  $(6.4\phantom{0}\pm0.4\phantom{0})\times 10^2$ 
& $\phantom{0}5.3 \pm 0.4 \pm 0.5  \pm 0.5 $ \\

$\decay{\Lb}{\Lz\Kp \pim}$ & $(6.18\pm0.32)\times 10^2$ &$\phantom{0}4.6 \pm 0.2 \pm 0.4 \pm 0.5 $ \\
$\decay{\Lb}{\Lz\Kp \Km}$ & $(1.92\pm0.05)\times 10^3$ & $10.7 \pm 0.3 \pm 0.4 \pm 1.1 $ \\
$\decay{\Xibz}{\Lz\pip \pim}$ & $(5.6\phantom{0}\pm2.7\phantom{0})\times 10^1$&$11.0 \pm 2.6 \pm 1.4  \pm 3.8 $ \\
$\decay{\Xibz}{\Lz\Km \pip}$ & $(1.19\pm0.15)\times 10^2$ & $10.4 \pm 1.4 \pm 1.2  \pm 3.5 $ \\[1pt]
$\decay{\Xibz}{\Lz\Kp \Km}$ & $(1.2\phantom{0}\pm0.9\phantom{0})\times 10^1$ &  $< 2.4$ $(2.8)$  at 90\% (95\%) CL\\ 
$\decay{\Lb}{\Lc(\to\Lz\pip) \pim}$ & $(5.25\pm0.07)\times 10^3$ &  ---\\ 
\hline
\end{tabular}
\end{table}

%%%%%%%%%%%%%% Branching fraction %%%%%%%%%%%%%%%%%%%%%%%%%%%%%
The branching-fraction ($\BF$) ratio of a signal decay to that of the control mode is  measured according to 
\begin{equation}\label{eq:br}
\frac{\BF\left(\decay{\Lb/\Xibz}{\Lz h^{+} h'^-}\right)}{\BF\left(\decay{\Lb}{ \Lc\left(\to\Lz \pip\right) \pim}\right)}	=
\frac{N_{\decay{\Lb/\Xibz}{\Lz h^{+} h'^{-}}}}{N_{\decay{\Lb}{\Lc\left(\to\Lz \pip\right) \pim}}}\times\frac {\epsilon_{\decay{\Lb}{\Lc\left(\to\Lz \pip\right) \pim}}}{ \epsilon_{\decay{\Lb/\Xibz}{\Lz h^{+} h'^{-}}} } \times \frac{f_{\Lb}}{f_{\Lb/\Xibz}}\,,
\end{equation}
where $N$ and $\epsilon$ are the yield and  efficiency for the considered decay, respectively, and the final factor is the ratio of $b$-quark fragmentation fractions~\cite{LHCb-PAPER-2014-004, LHCb-PAPER-2018-047}. 
The yields are determined 
through the fit to data while 
the efficiencies are determined from simulation. In the simulation the $\pt$ and rapidity distributions of the $\Lb$ baryon~\cite{LHCb-PAPER-2015-037}, as well as the Dalitz plot of the $\Lb/\Xibz$ decays, are corrected
to match those in data. 
The efficiencies are at the level of $10^{-4}$, with the efficiency ratio in the range 0.8--2.9 depending on the signal channel. 
For $\Xibz$ decays, due to the limited data sample, the $\pt$ and rapidity are not corrected and a $10\%$ systematic uncertainty is assigned to the efficiency. 

The branching-fraction results are summarized in Table~\ref{tab:branching_ratios}, where the uncertainties are statistical, systematic and due to the uncertainty of the control channel branching fraction~\cite{BESIII:2015bjk, LHCb-PAPER-2014-004}.
As no significant contribution from the \XiKK decay is found, upper limits are determined on its branching fraction at 90\% and 95\% confidence levels (CL), by integrating the positive side of the profile likelihood~\cite{LHCb-PAPER-2022-004}. 

%%%%%%%%%%%%%%%%%%%%%%% \CP Violation %%%%%%%%%%%%%%%%%%%%%%%%%%%%%%%%%

Four channels with sufficiently high yields, including three $\Lb$ decay modes and the \XiPK decay mode, are selected for further investigation of \CP violation.
The \CP asymmetry of the decay to a final state $f$ is defined as
\begin{equation}
\ACP(\decay{\Lb/\Xibz}{f})\equiv \frac{\Gamma(\decay{\Lb/\Xibz}{f}) - \Gamma(\decay{\Lbbar/\Xibbarz}{\overline{f}})}{\Gamma(\decay{\Lb/\Xibz}{f}) + \Gamma(\decay{\Lbbar/\Xibbarz}{\overline{f}})},
\end{equation}
where $\Gamma$ is the partial decay rate defined without inclusion of its charge-conjugate process. 
The raw asymmetry of signal yields between baryon and antibaryon decays, denoted as $\ACPraw$, is first extracted directly from the mass fits. This is then corrected to account for two factors: the asymmetry of the baryon and antibaryon production rates, $A_\text{P}$, and the asymmetry of the final-state detection and selection efficiencies, $A_\text{exp}$. 
To reduce systematic uncertainties, the difference between the $\CP$ asymmetry of each signal decay and the $\decay{\Lb}{\Lc(\Lz\pip)\pim}$ decay, $\Delta\ACP$, is measured. 
Assuming there is no \CP violation for the control mode, valid within the experimental uncertainties of this analysis,
$\Delta\ACP$ gives the measurement of the $\CP$ asymmetry for the signal decay.

\begin{figure}[!t]
    \centering
    \includegraphics[width=0.48\textwidth]{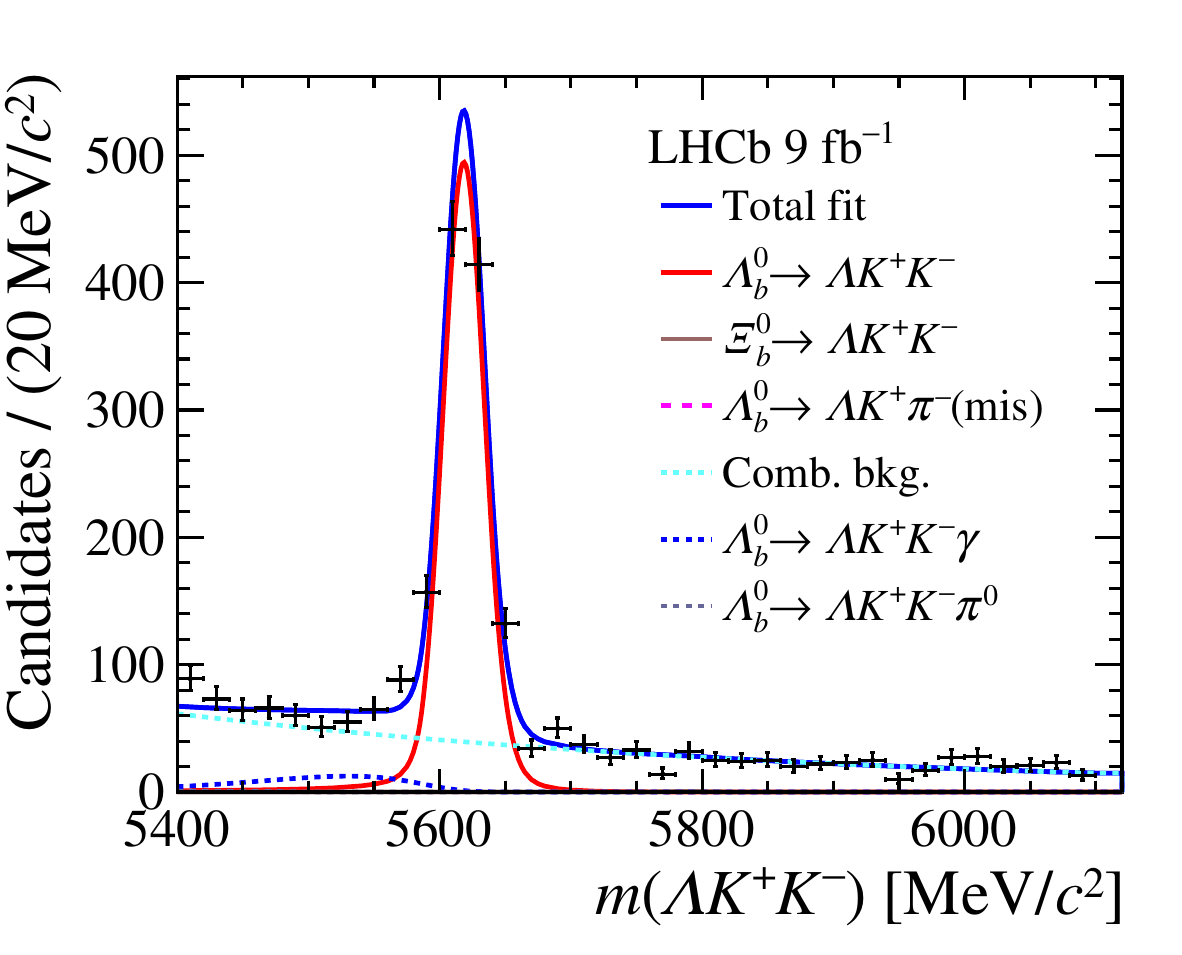}
    \put(-170,140){(a)}
    \includegraphics[width=0.48\columnwidth]{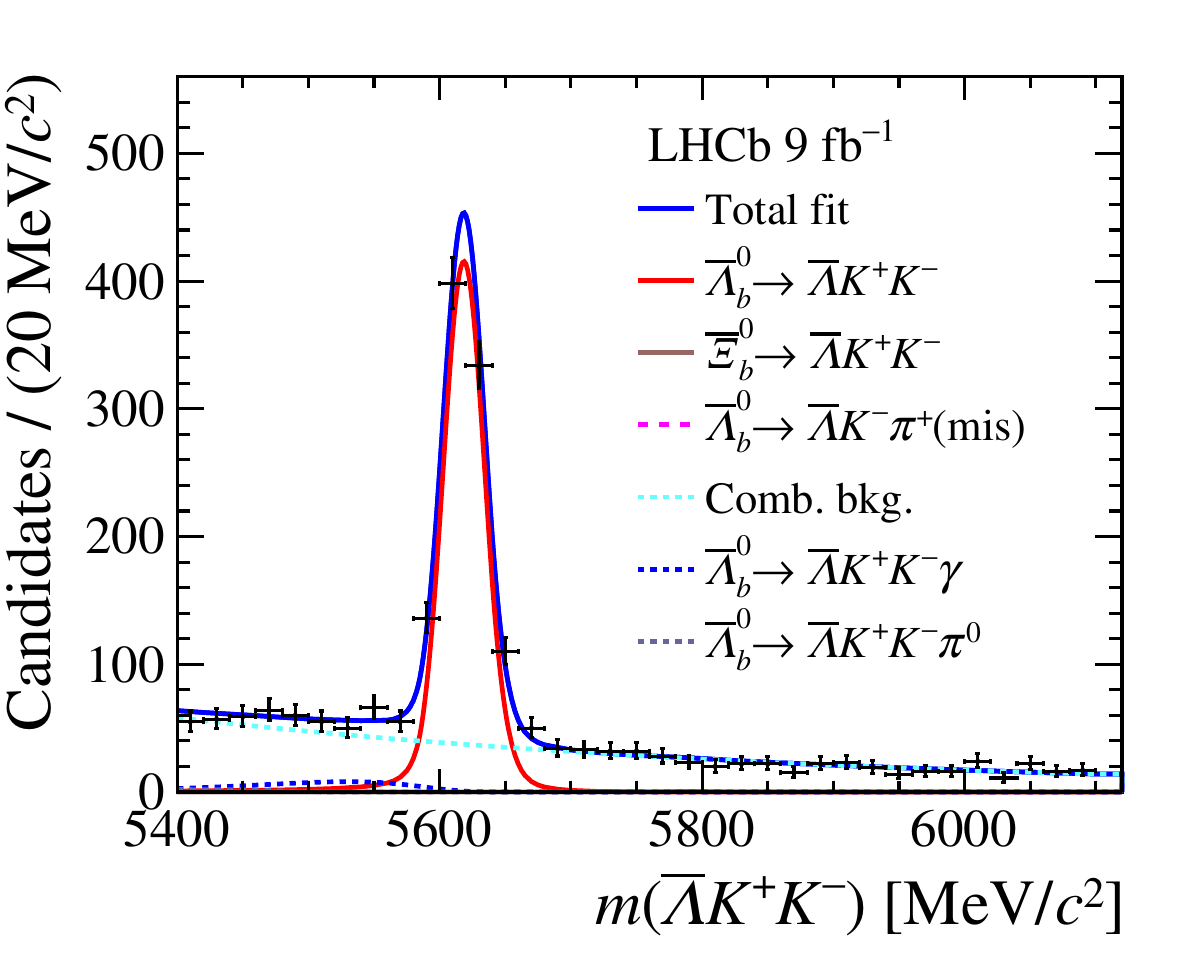}
    \put(-170,140){(b)}
\caption{Mass distributions of (a) $\decay{\Lb}{\Lz K^+K^-}$ and (b) $\decay{\Lbbar}{\Lbar \Kp\Km}$
    decays, with the fit projections.
    }\label{fg:UDLb}
\end{figure}

The $\Lb$ production asymmetries in $pp$ collisions at $\sqs=7\tev$ and $8\tev$ have previously been measured at~\lhcb~\cite{LHCb-PAPER-2018-025}, but there is no equivalent measurement yet at $\sqs=13\tev$.
As the $\Lb$ production asymmetry is expected to be smaller at higher energies and mostly cancels between the signal and control channel, 
the $A_\text{P}$ measured for $\sqs=8\tev$ is used for the $\Delta \ACP(\Lb)$ measurements at $\sqs=13\tev$. 
Assuming isospin symmetry between the $\Xibz$ and $\Xibm$ cross-sections in $pp$ collisions,  the $\Xibz$ production asymmetry is taken to be the same as that of the $\Xibm$ baryon, which has been measured by the \lhcb experiment~\cite{LHCb-PAPER-2018-047}.
The detection asymmetry encompasses the asymmetries in the final-state reconstruction, the trigger selection and the PID selection.
The reconstruction asymmetries for pions, kaons and protons have been measured as a function of particle momenta using control samples of 
$\decay{\Dp}{\KS\pip}$, $\decay{\Dp}{\Km\pip\pip}$,  
$\decay{\Dstarp}{\Dz(\to \Km\pip\pip\pim)\pip}$ decays
~\cite{LHCb-PAPER-2014-013}, 
and simulated samples of $\decay{\Lb}{\Lc(\to p\Km\pip)\mun\neumb}$ decays~\cite{LHCb-PAPER-2018-025}. 
The detection asymmetry for each final-state particle is then weighted by its momentum distribution in the signal and control modes to get an averaged result, accounting for the kinematics of both modes. 
The PID and trigger selection asymmetries are obtained in a similar way using data~\cite{LHCb-PUB-2016-021,LHCb-DP-2019-001}. The largest detection asymmetry, due to proton reconstruction, mostly cancels  between the signal and control modes. 
These correction terms $\Delta A_\text{P}$ and $\Delta A_\text{exp}$ are shown in Table~\ref{tab:correction_terms} in the End Matter, and are all consistent with zero for $\Lb$ decays with uncertainties around 0.002 and 0.010, respectively.
The $\Delta\ACP$ quantities, integrated over the phase space, are measured for the four decays to be
\[
\begin{aligned}
 \Delta\ACP\left(\Lb \to \Lz\pip\pim\right)\ \  & = -0.013 \pm 0.053 \pm 0.018, \\
 \Delta\ACP\left(\Lb \to \Lz\Kp\pim\right)\    & = -0.118 \pm 0.045 \pm 0.021, \\
 \Delta\ACP\left(\Lb \to \Lz\Kp\Km\right)     & = \phantom{-}0.083 \pm 0.023 \pm 0.016, \\
 \Delta\ACP\left(\Xibz \to \Lz\Km\pip\right)\    & = \phantom{-}0.27\phantom{0} \pm 0.12\phantom{0} \pm 0.05, \\
\end{aligned}
\]
where the first uncertainties are statistical and the second are systematic.
The $\Delta\ACP$ measurement for the \LzKK decay has a significance of  $3.1\,\sigma$ based on the negative log-likelihood method~\cite{LHCb-PAPER-2023-018}, accounting for both statistical and systematic uncertainties. This significance is confirmed by using ensembles of pseudoexperiments. 

\begin{figure}[!t]
	\centering
	\begin{minipage}[t]{0.48\columnwidth}
		\centering
		\includegraphics[width=1\columnwidth]{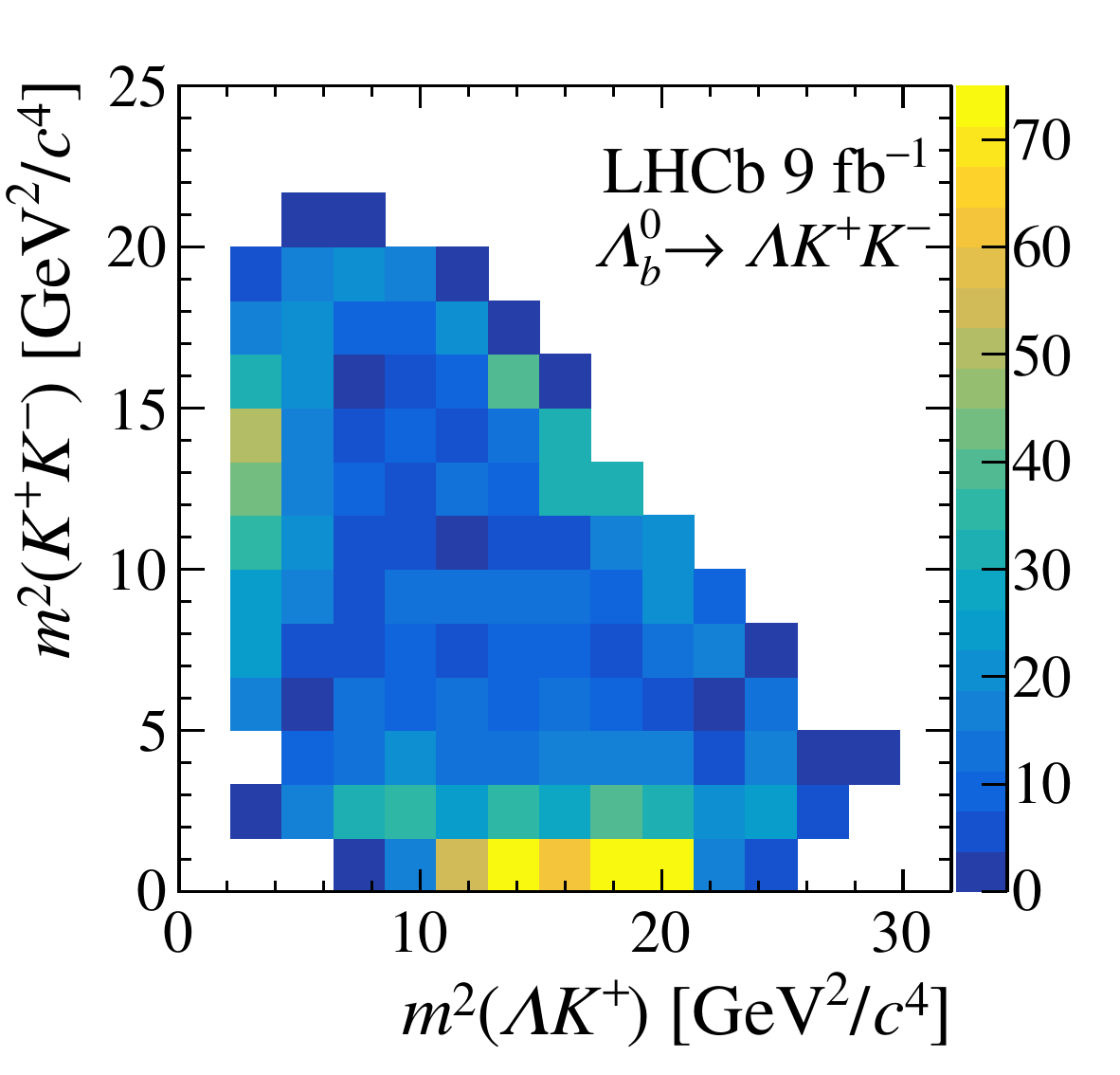} 
  \put(-173,175){(a)}
	\end{minipage}%
	\begin{minipage}[t]{0.48\columnwidth}
		\centering
		\includegraphics[width=\columnwidth]{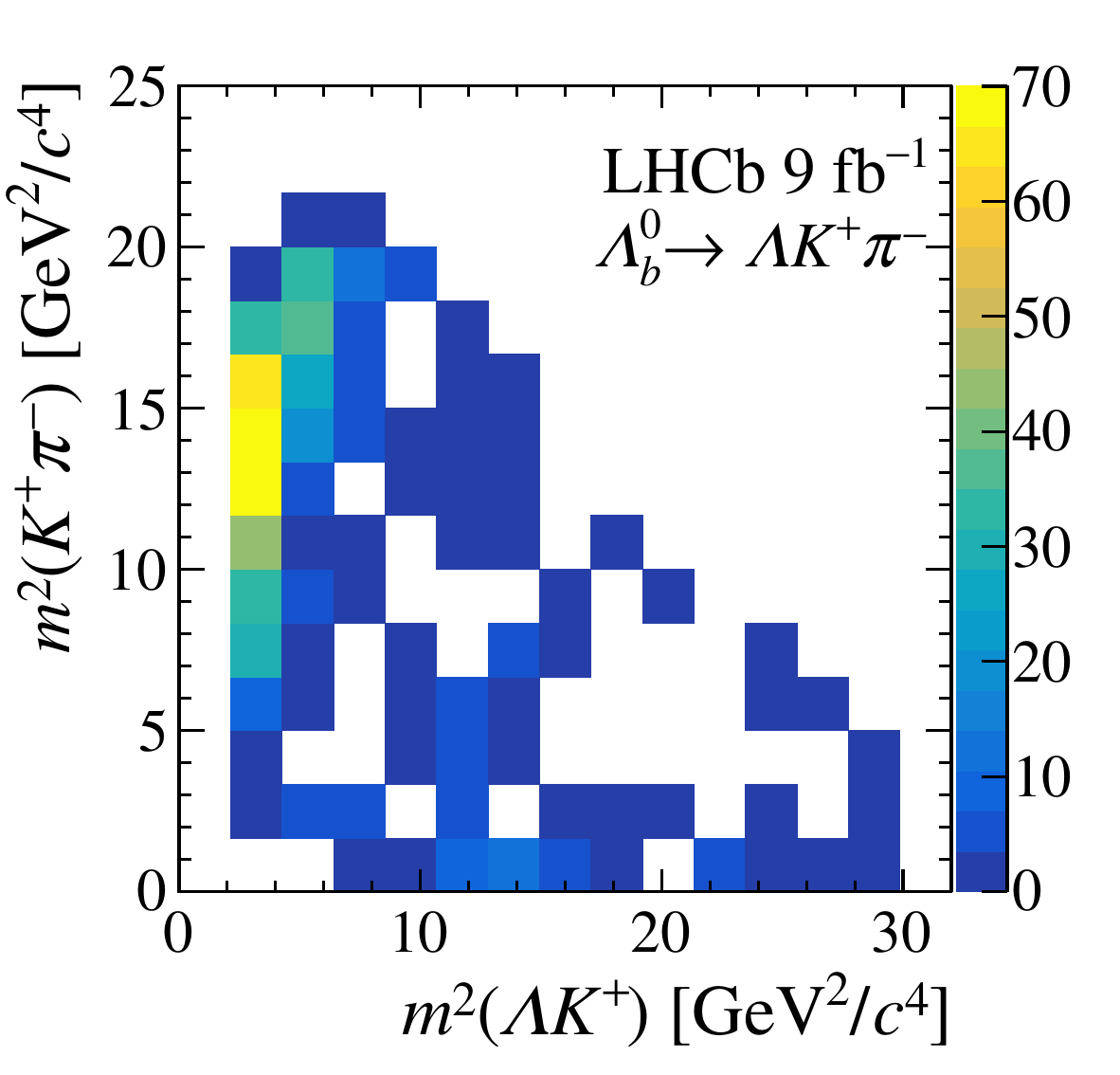}
  \put(-173,175){(b)}
	\end{minipage}
	\begin{minipage}[t]{0.48\columnwidth}
		\centering
		\includegraphics[width=1\columnwidth]{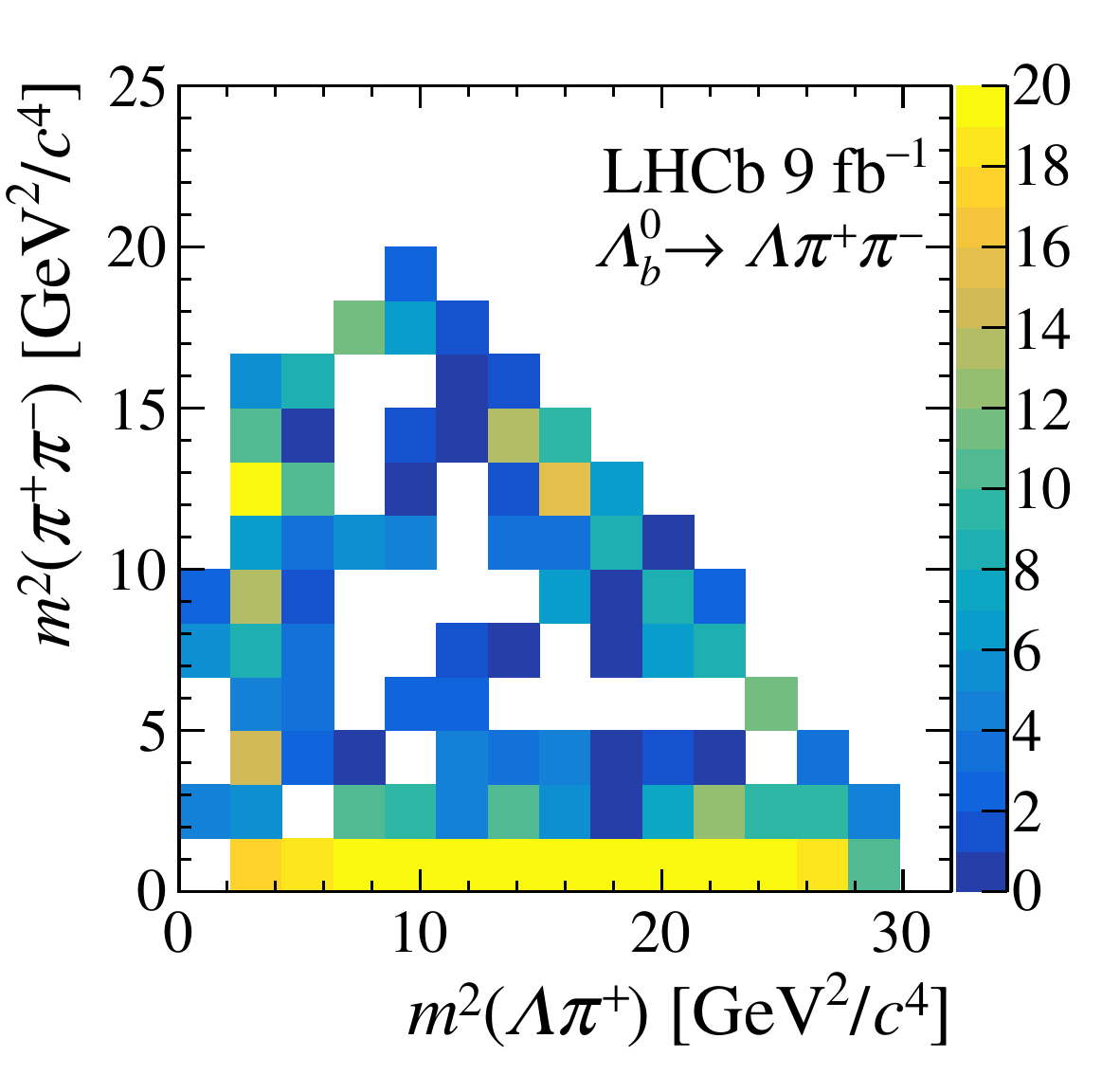}
  \put(-173,175){(c)}
	\end{minipage}
	\caption{Dalitz plots of (a) $\decay{\Lb}{\Lz K^+K^-}$, (b) $\decay{\Lb}{\Lz K^+\pi^-}$, (c) $\decay{\Lb}{\Lz \pi^+\pi^-}$ decays. Background contributions are subtracted using the \sPlot technique. The coordinates are calculated after a kinematic fit which constrains the \Lb and \Lz baryon masses to their known values~\cite{PDG2024}.}\label{fg:DalitzLb}
\end{figure}

\begin{figure}[!t]
\centering
\includegraphics[width=0.48\columnwidth]{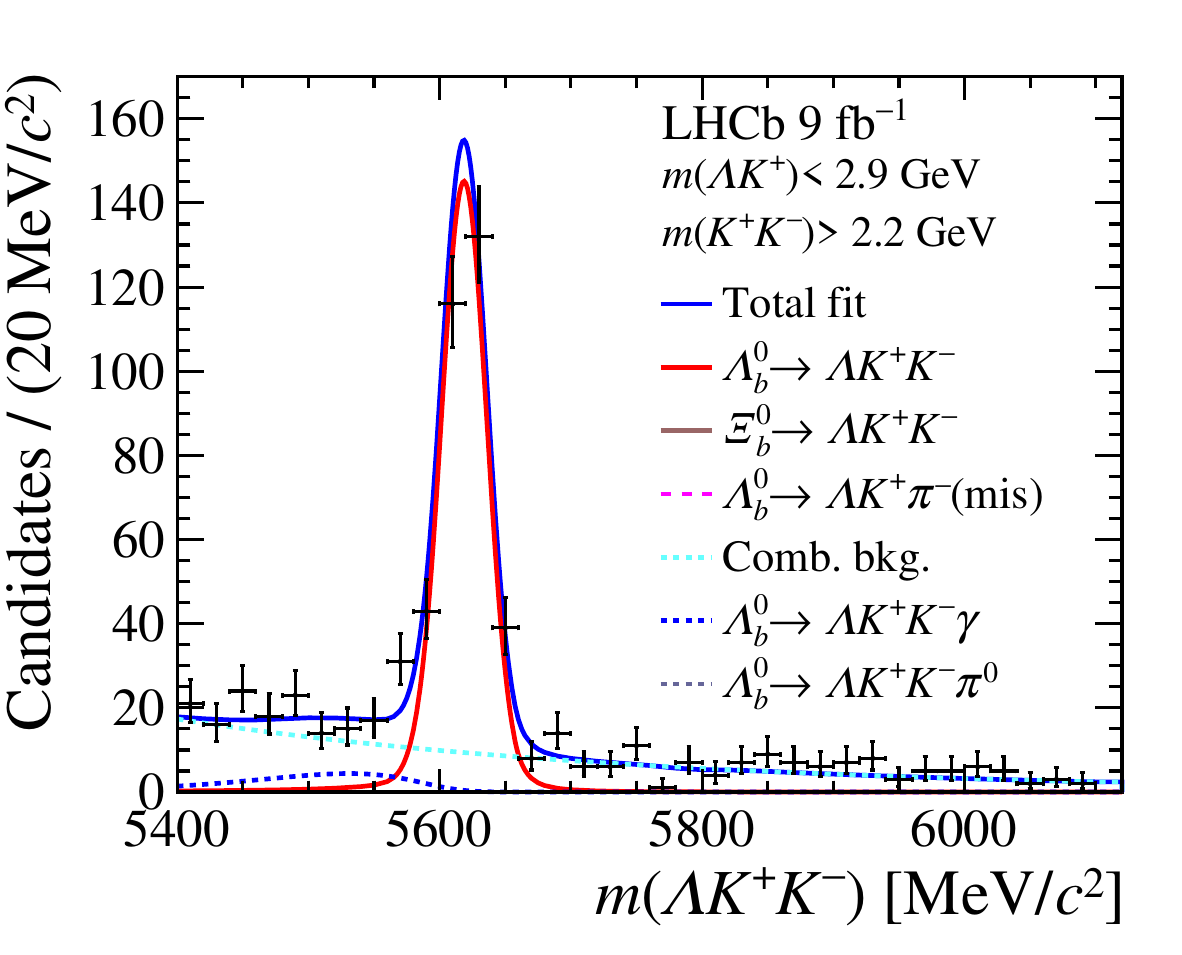}
\put(-170,140){(a)}
\includegraphics[width=0.48\columnwidth]{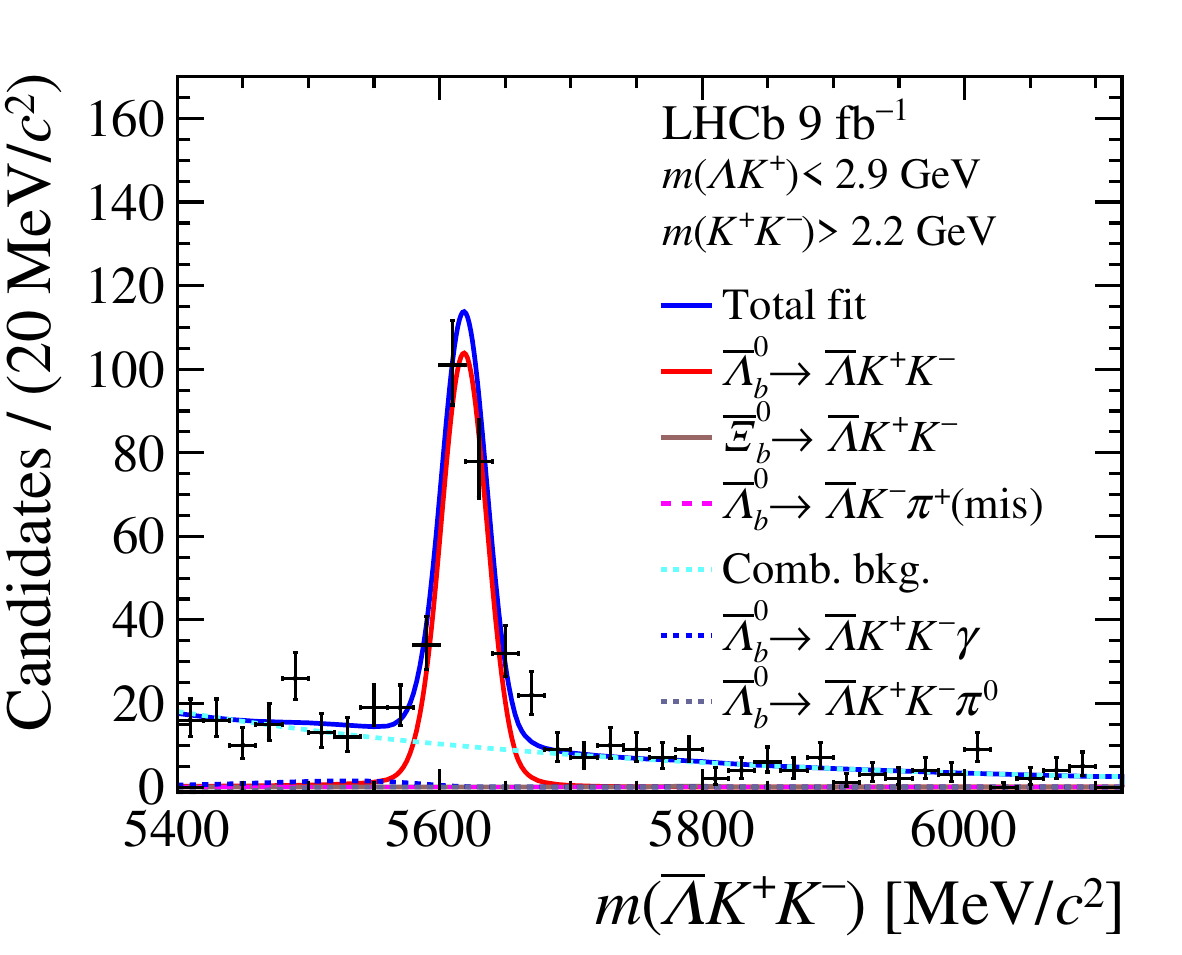}
\put(-170,140){(b)}
\caption{Mass distributions of (a) $\decay{\Lb}{\Lz K^+K^-}$ and (b) $\decay{\Lbbar}{\Lbar \Kp\Km}$
    decays in $N^*$ resonance-dominated regions. Also shown are the fit results.}\label{fg:UDresLb}
\end{figure}

The mass distributions of \LzKK for both baryon and antibaryon decays, with fit results also plotted, are shown in Fig.~\ref{fg:UDLb} where a clear difference in signal yields between $\Lb$ and $\Lbbar$ decays can be seen.  
The decay is dominated by intermediate $N^{*+}(\to \Lz \Kp)$ or $\phi(\to\Kp\Km)$ resonances,  as can be seen in  the \LzKK  Dalitz plot of Fig.~\ref{fg:DalitzLb}~(a), where background 
contributions are subtracted using the \sPlot technique~\cite{Pivk:2004ty}.
To investigate whether these resonances are the source of the \CP asymmetry, separate $\Delta\ACP$ measurements are performed within these two resonance-dominated regions.
In the region dominated by the $N^{*+}$ resonance, 
 the
asymmetry 
 is determined to be 
\mbox{$\Delta\ACP=0.165 \pm 0.048\pm 0.017$}, which differs from zero by $3.2\,\sigma$. 
The mass distributions of the \LzKK and $\decay{\Lbbar}{\Lbar \Kp\Km}$ decays and their fit projections, within the region, are shown in Fig.~\ref{fg:UDresLb}, demonstrating the difference 
between $\Lb$ and $\Lbbar$ yields.
The \CP asymmetry in the $\phi$ region is consistent with zero. 
A possible variation of the \CP asymmetry across the Dalitz plot is also studied in 10 equally populated  Dalitz bins defined using an adaptive binning scheme. 
The results are consistent with \CP symmetry.

The significances for \CP violation in
 \LzKP, \LzPP and \XiPK decays are $2.4\,\sigma$, $0.2\,\sigma$ and $2.1\,\sigma$, respectively. Further searches for \CP violation are also performed for the two $\Lb$ decays
   both in resonance-dominated regions  (see Fig.~\ref{fg:DalitzLb}~(b, c) and Table~\ref{ta:resonanceM}) and with an adaptive binning scheme. The results are all consistent with \CP symmetry.
For the $\Xibz$ decays, no localized \CP asymmetry searches are performed due to the low signal yields.
   
\begin{table}[!t]
	\caption{Definitions of the resonance-dominated regions and the corresponding  $\Delta\ACP$ values. The symbol $f$ represents multiple resonances at low $\pip\pim$ mass.} \label{ta:resonanceM}
	\centering
 {\renewcommand{\arraystretch}{1.1} 
	\begin{tabular}{lccc}
		\hline Channel & $m(h^+h'^-)$ & $m(\Lz h^+)$ & $\Delta\ACP$\\
		\hline
		$\decay{\Lb}{\Lz	\phi\left(\to\Kp\Km\right)}$	&  $<1.10 \gevcc$  & 
        --- &$\phantom{-}0.150 				\pm	0.055 	\pm	0.021$ 	\\
		$\decay{\Lb}{N^{*+}(\to\Lz	\Kp)\Km}$    & $>2.20 \gevcc$	& $<2.90\gevcc$ & $\phantom{-}0.165 				\pm	0.048 	\pm	0.017$ \\
  
		$\decay{\Lb}{N^{*+}(\to\Lz \Kp)\pim}$		& --- &$< 2.30 \gevcc$	& $-0.078 				\pm	0.051 	\pm	0.027$ \\
		$\decay{\Lb}{\Lz f\left(\to\pip\pim\right)}$		& $<1.70 \gevcc$ &---	& $\phantom{-} 				0.088\pm	0.069 	\pm	0.021$ \\
		\hline 
	\end{tabular}
 }
\end{table}

Cross-checks are performed to investigate the stability of the branching fraction and $\Delta\ACP$ measurements. 
For the global asymmetries and branching fractions, results are obtained in different data-taking periods, as well as with different magnet polarities, and are found to be consistent. 
For the measurements in different  resonance-dominated regions,  alternative definitions of the mass regions are used, and similar results as the nominal ones are obtained. 

%%%%%%%%%% systematics %%%%%%%%%%%%%%%
Various sources of systematic uncertainties on the branching fraction and $\Delta\ACP$ measurements are considered. 
The uncertainty due to the imperfect modelling of the mass distributions is evaluated by using alternative models for each component, including an Hypatia function~\cite{Santos:2013gra} for the signal model and a second-order polynomial function for the combinatorial background. 
For the $\Delta\ACP$ measurements, an additional uncertainty arises from using shared fit parameters for baryon and antibaryon decays. This is assessed by removing this constraint and assigning the resulting $\Delta\ACP$ shifts as systematic uncertainties. 
The systematic uncertainty from the efficiency ratio has several contributions. 
The first contribution arises from the finite size of simulation samples, which is propagated to the branching fraction and $\Delta\ACP$ measurements using pseudoexperiments. 
Another contribution is due to the robustness of efficiency corrections, which are studied in alternative scenarios. For example, the effect of the vetoing of charm hadrons is studied by varying the vetoed mass regions, a new efficiency map is obtained to calculate the corresponding branching fraction and the difference is taken as a systematic uncertainty. 
The uncertainties on the production and experimental asymmetries are propagated to the $\Delta\ACP$ measurements using pseudoexperiments and 
largely cancel
in the difference of signal and control mode asymmetries. 
The total systematic uncertainties are obtained by summing all contributions in quadrature.

%%%%%%%%%%%% conclusion %%%%%%%%%%%%%%%%%

In summary, $\decay{\Lb/\Xibz}{\Lz h^+h'^-}$ decays are studied using $pp$ collision data collected by the \lhcb experiment during \lhc Runs 1--2. 
The \LzPP and \XiPK decays are observed for the first time, and evidence is also found for the \XiPP decay. The branching-fraction measurements of $\decay{\Lb/\Xibz}{\Lz h^+h'^-}$ decays are more precise than and
supersede previous \lhcb results~\cite{LHCb-PAPER-2016-004}. 
The $\CP$ asymmetries are measured for $\decay{\Lb}{\Lz h^+h'^-}$ and $\decay{\Xibz}{\Lz \Km\pip}$ decays, with respect to the $\Lb\to\Lc(\to\Lz\pip)\pim$ decay. 
Evidence for $\CP$ violation is found in the \LzKK decay for the first time, with $\Delta\ACP = (8.3\pm2.8)\%$ integrated over the final-state phase space. The \CP asymmetry is enhanced in the $N^{*+}$ mass region, where it is measured to be $\Delta\ACP=(16.5\pm5.1)\%$. No evidence of \CP violation is found for other \Lb/\Xibz decays studied.
These measurements represent an important step towards establishing \CP violation in baryon decays, setting the stage for future studies of quasi-two-body decays. 

%%%% acknowledgements %%%%%

\section*{Acknowledgements}
%
% These Acknowledgements valid from 3-May-2019
%
\noindent We express our gratitude to our colleagues in the CERN
accelerator departments for the excellent performance of the LHC. We
thank the technical and administrative staff at the LHCb
institutes.
We acknowledge support from CERN and from the national agencies:
CAPES, CNPq, FAPERJ and FINEP (Brazil); 
MOST and NSFC (China); 
CNRS/IN2P3 (France); 
BMBF, DFG and MPG (Germany); 
INFN (Italy); 
NWO (Netherlands); 
MNiSW and NCN (Poland); 
MCID/IFA (Romania); 
%MSHE (Russia); 
MICIU and AEI (Spain);
SNSF and SER (Switzerland); 
NASU (Ukraine); 
STFC (United Kingdom); 
DOE NP and NSF (USA).
We acknowledge the computing resources that are provided by CERN, IN2P3
(France), KIT and DESY (Germany), INFN (Italy), SURF (Netherlands),
PIC (Spain), GridPP (United Kingdom), 
%RRCKI and Yandex LLC (Russia), 
CSCS (Switzerland), IFIN-HH (Romania), CBPF (Brazil),
and Polish WLCG (Poland).
We are indebted to the communities behind the multiple open-source
software packages on which we depend.
Individual groups or members have received support from
ARC and ARDC (Australia);
Key Research Program of Frontier Sciences of CAS, CAS PIFI, CAS CCEPP, 
Fundamental Research Funds for the Central Universities, 
and Sci. \& Tech. Program of Guangzhou (China);
Minciencias (Colombia);
EPLANET, Marie Sk\l{}odowska-Curie Actions, ERC and NextGenerationEU (European Union);
A*MIDEX, ANR, IPhU and Labex P2IO, and R\'{e}gion Auvergne-Rh\^{o}ne-Alpes (France);
%RFBR, RSF and Yandex LLC (Russia);
AvH Foundation (Germany);
ICSC (Italy); 
%GVA, XuntaGal, GENCAT, Inditex, InTalent and Prog.~Atracci\'on Talento, CM (Spain);
Severo Ochoa and Mar\'ia de Maeztu Units of Excellence, GVA, XuntaGal, GENCAT, InTalent-Inditex and Prog. ~Atracci\'on Talento CM (Spain);
SRC (Sweden);
the Leverhulme Trust, the Royal Society
 and UKRI (United Kingdom).

\clearpage

%%%% supplementary %%%%%%
\newpage
\section*{End Matter}
\label{sec:Supplementary}

\section{Summary figures and tables for adaptive binnning scheme}
\begin{figure}[h]
    \centering
    \begin{minipage}[h]{0.33\columnwidth}
        \centering
        \includegraphics[width=\columnwidth]{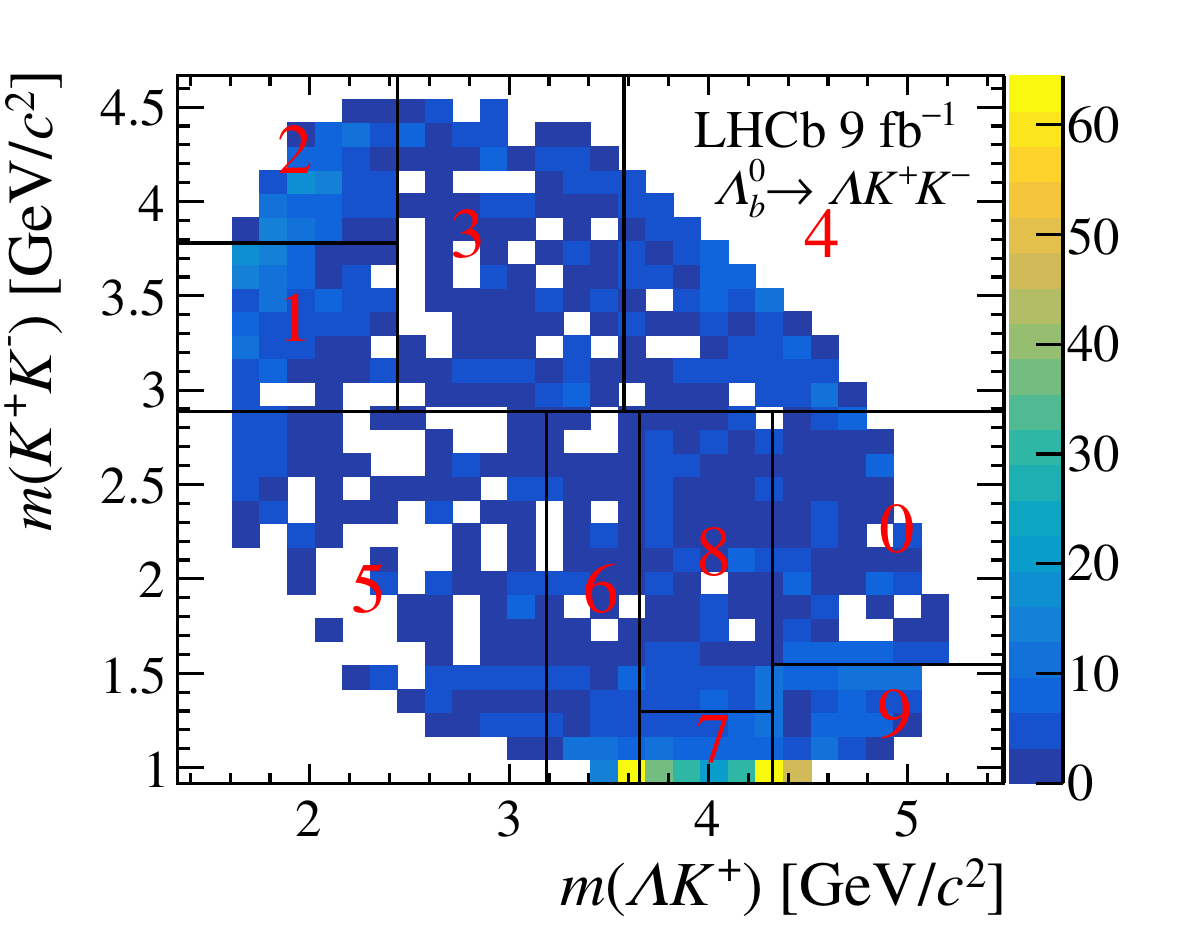}
    \end{minipage}\put(-75, -75){(a)}
    \begin{minipage}[h]{0.33\columnwidth}
        \centering
        \includegraphics[width=\columnwidth]{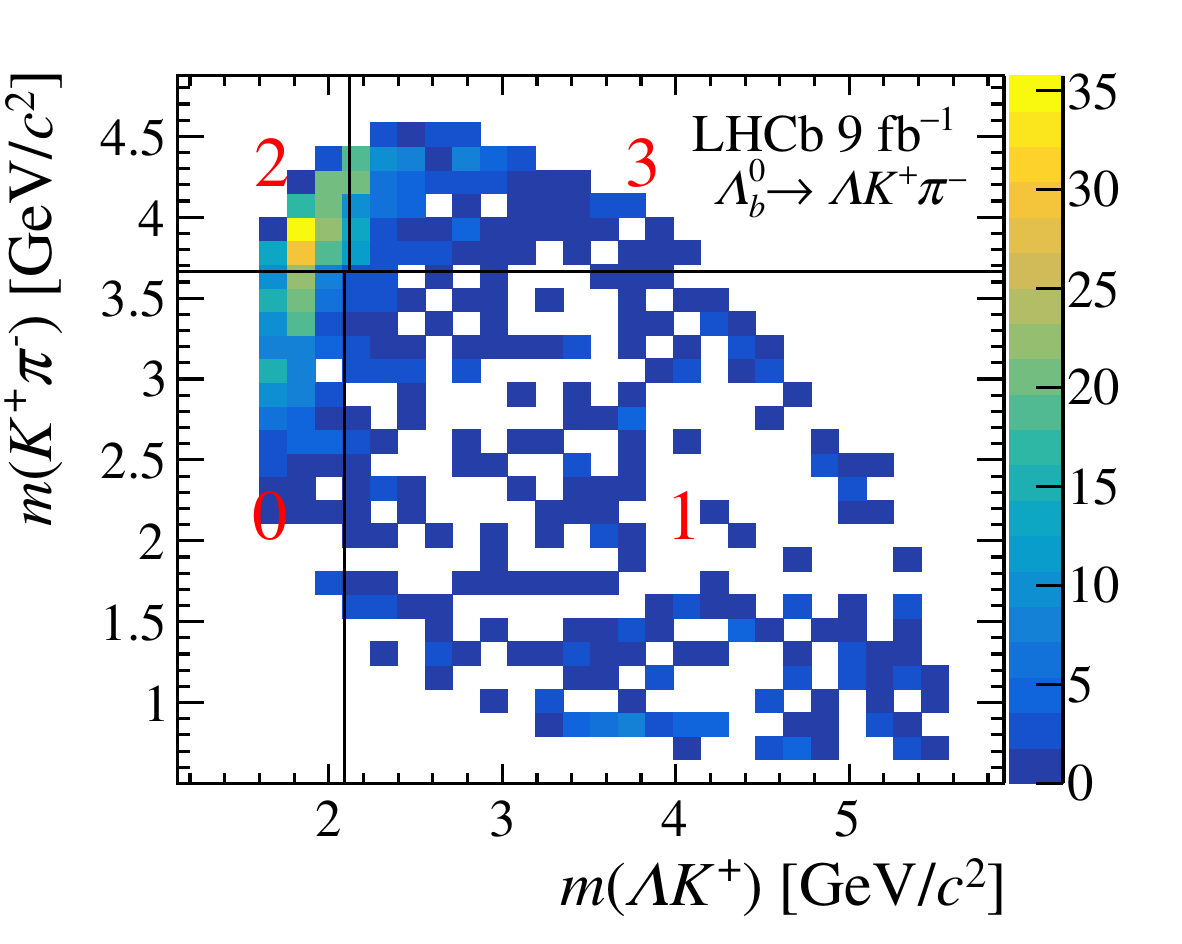}
    \end{minipage}\put(-75, -75){(b)}
    \begin{minipage}[h]{0.33\columnwidth}
        \centering
        \includegraphics[width=\columnwidth]{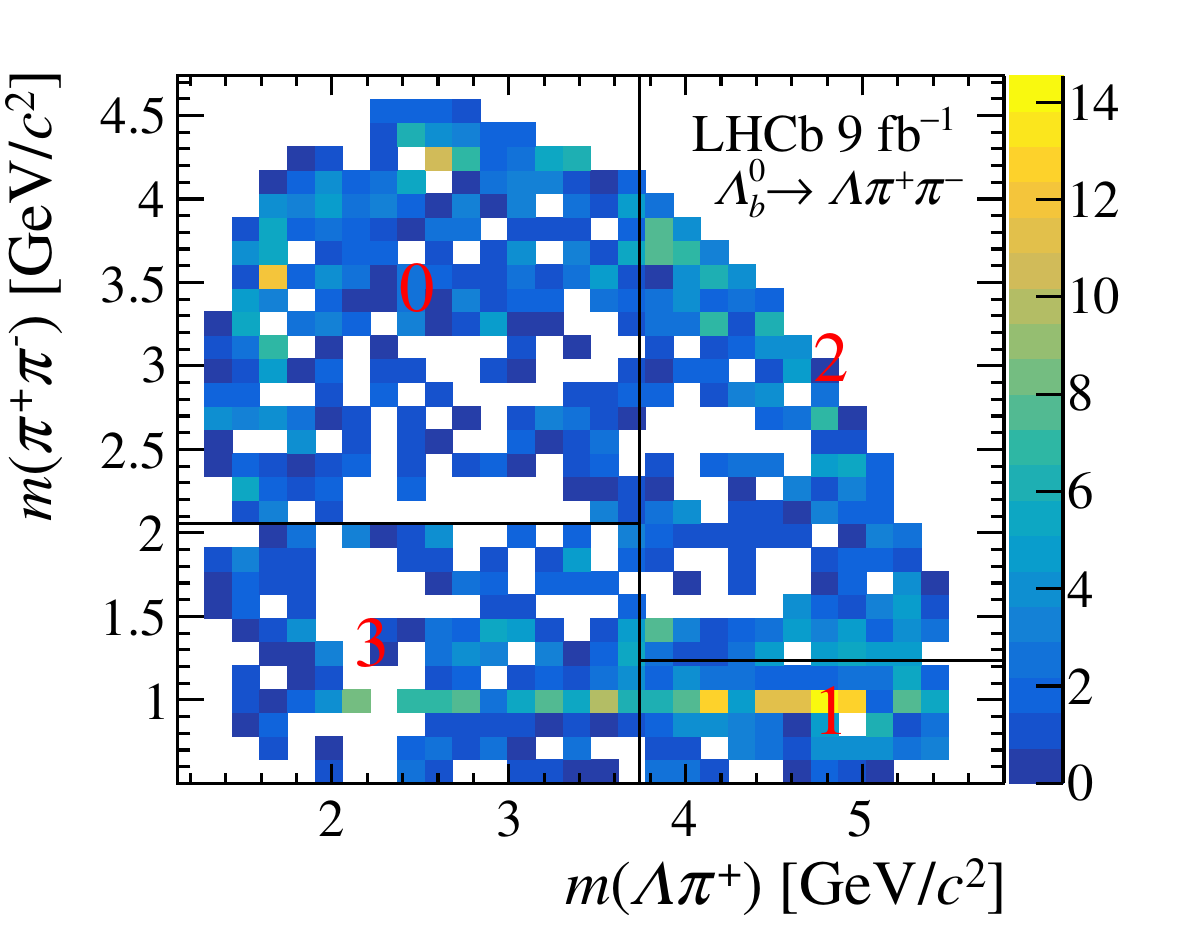}
    \end{minipage}\put(-75, -75){(c)}\\
\caption{Two dimensional mass distributions for (a) $\decay{\Lb}{\Lz K^+K^-}$, (b) $\decay{\Lb}{\Lz K^+\pi^-}$, and (c) $\decay{\Lb}{\Lz \pi^+\pi^-}$ 
    decays in data. The boundaries for the adaptive binning scheme are drawn as solid lines.}\label{fig:binning}
\end{figure}

Figure~\ref{fig:binning} shows the two dimensional mass distributions for (a) $\Lb\to\Lz\Kp\Km$, (b) $\Lb\to\Lz\Kp\pim$, and (c) $\Lb\to \Lz\pip\pim$, along with the bin boundaries used for the adaptive binning scheme.
Table~\ref{ta:binning} lists the bin definitions used for each decay mode in the adaptive binning scheme and the per-bin \CP asymmetry measurements.

 \begin{table}[!b]
	\centering
{\renewcommand{\arraystretch}{1.1} 	\caption{Boundaries of the adaptive binning scheme and the $\Delta\ACP$ measurements from each bin, the first uncertainty is statistical and the second is systematic.  The variables of the $x$ and $y$ axes and the bin numbers in the table are those presented in Fig.~\ref{fig:binning}. The reported ranges are expressed in \gevcc.} \label{ta:binning}

	\begin{tabular}{lcccccc}
\hline
Channel			& bin number &	$x$-low	&	$x$-high	&	$y$-low	&	$y$-high	& $\Delta\ACP$\\
\hline
\multirow{4}{*}{	$\Lb\rightarrow\Lz\pip\pim$}	& 0	&	1.13	&	3.74	&	2.05	&	4.74	&  $-0.483 		    \pm	0.200 	\pm	0.043$ \\	
			&    1 &	3.74	&	5.50	&	0.50	&	1.24	&    $\phantom{-}0.147 	\pm	0.092 \pm	0.026$  \\	
			&    2 &	3.74	&	5.50	&	1.24	&	4.74	&    $\phantom{-}0.058 	\pm	0.114 	\pm	0.028$\\	
			&    3 &	1.13	&	3.74	&	0.50	&	2.05	&    $\phantom{-}0.067 	\pm	0.111 	\pm	0.028$\\	\hline
\multirow{4}{*}{	$\Lb\rightarrow\Lz	\Kp\pim$}	& 0 &	1.13	&	2.09	&	0.50	&	3.66	&$ -0.153 				\pm	0.079 	\pm	0.027$ \\	
			&    1 &	2.09	&	5.49	&	0.50	&	3.66	&   $-0.284 	\pm	0.188 	\pm	0.041$ \\	
			&    2 &	1.13	&	2.12	&	3.66	&	4.87	&   $-0.006 	\pm	0.062 	\pm	0.028$ \\	
			&    3 &	2.12	&	5.49	&	3.66	&	4.87	&   $-0.264 	\pm	0.125 	\pm	0.030$ \\	\hline
\multirow{10}{*}{	$\Lb\rightarrow\Lz	\Kp\Km$}	& 0 &	4.32	&	5.08	&	1.55	&	2.88	& $\phantom{-}0.017 				\pm	0.092 	\pm	0.025$    \\	
			&    1 &	1.33	&	2.44	&	2.88	&	3.78  &   $\phantom{-}0.188 	\pm	0.075 	\pm	0.023$	    \\	
			&    2 &	1.33	&	2.44	&	3.78	&	4.67  &   $\phantom{-}0.062 	\pm	0.077 	\pm	0.022$	    \\	
			&    3 &	2.44	&	3.58	&	2.88	&	4.67  &    $\phantom{-}0.064 				\pm	0.093 	\pm	0.024$    \\	
			&    4 &	3.58	&	5.08	&	2.88	&	4.67  &    $\phantom{-}0.088 				\pm	0.077 	\pm	0.022$    \\	
			&    5 &	1.33	&	3.19	&	0.92	&	2.88  &    $\phantom{-}0.061 	\pm	0.089 	\pm	0.024$  \\	
			&    6 &	3.19	&	3.66	&	0.92	&	2.88  &    $\phantom{-}0.066 				\pm	0.088 	\pm	0.024$  \\	
			&    7 &	3.66	&	4.32	&	0.92	&	1.30  &    $\phantom{-}0.168 				\pm	0.070 	\pm	0.021$\\	
			&    8 &	3.66	&	4.32	&	1.30	&	2.88  &   $-0.002 				\pm	0.080 	\pm	0.023$ \\	
			&    9 &	4.32	&	5.08	&	0.92	&	1.55  &    $\phantom{-}0.025 				\pm	0.074 	\pm	0.022$\\

		\hline  
	\end{tabular}
 }
\end{table}

\section{Summary of the fit results}

\begin{figure}[!b]
  \begin{center}
    \includegraphics[width=0.35\columnwidth]{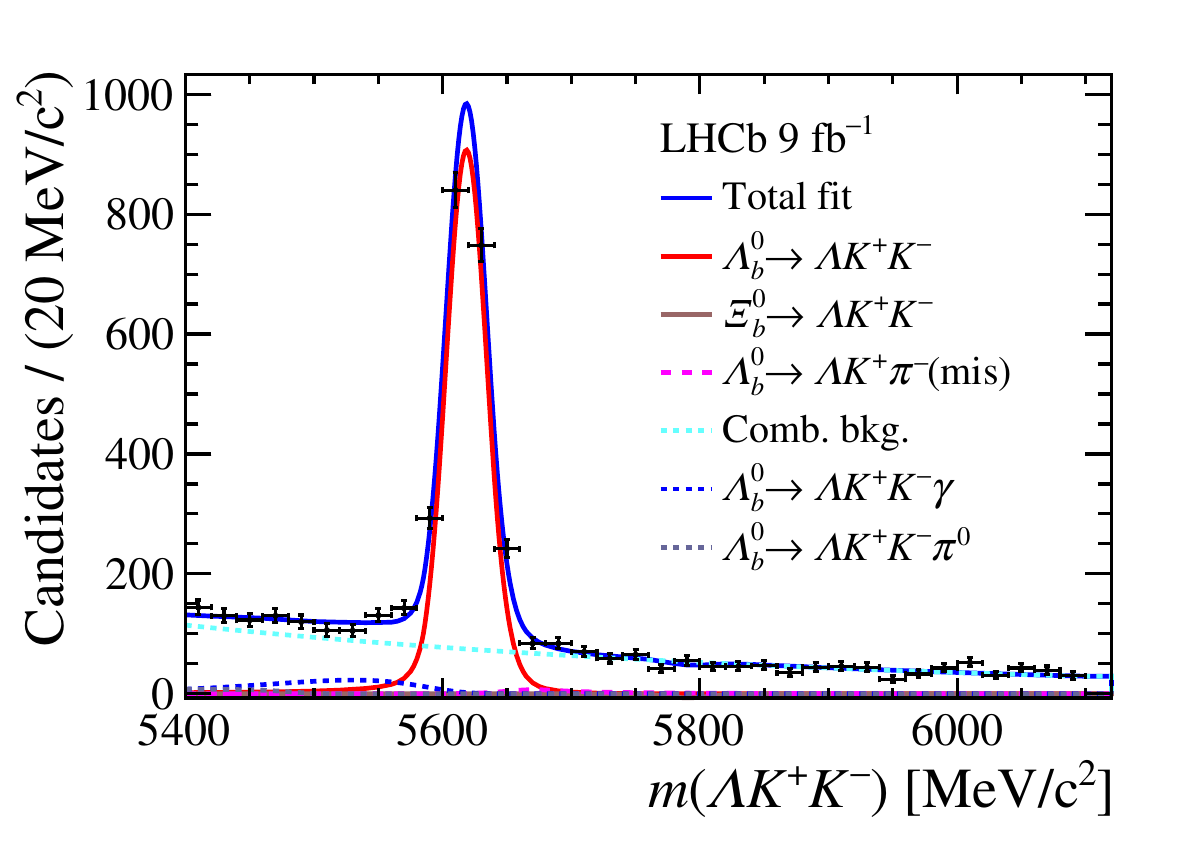}\put(-125,88){(a)}
    \includegraphics[width=0.35\columnwidth]{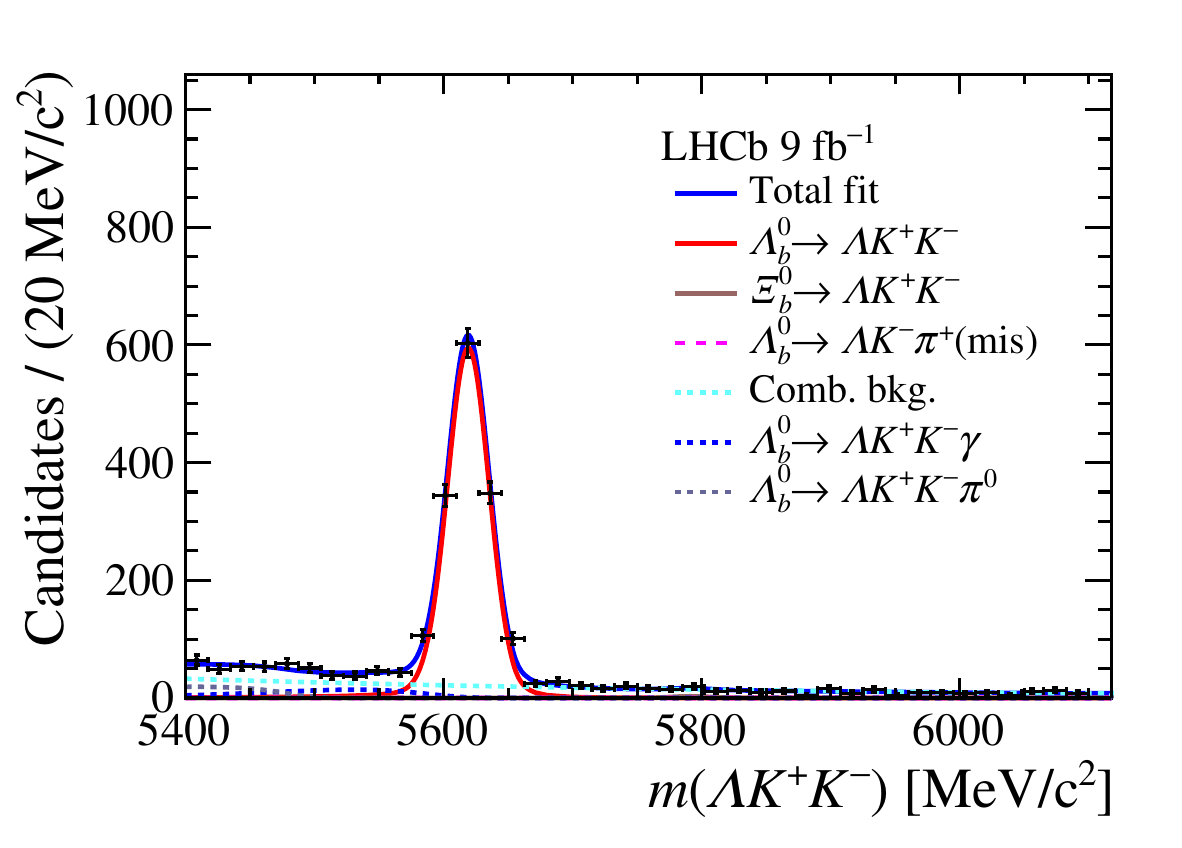}\put(-125,88){(b)}\\
    \includegraphics[width=0.35\columnwidth]{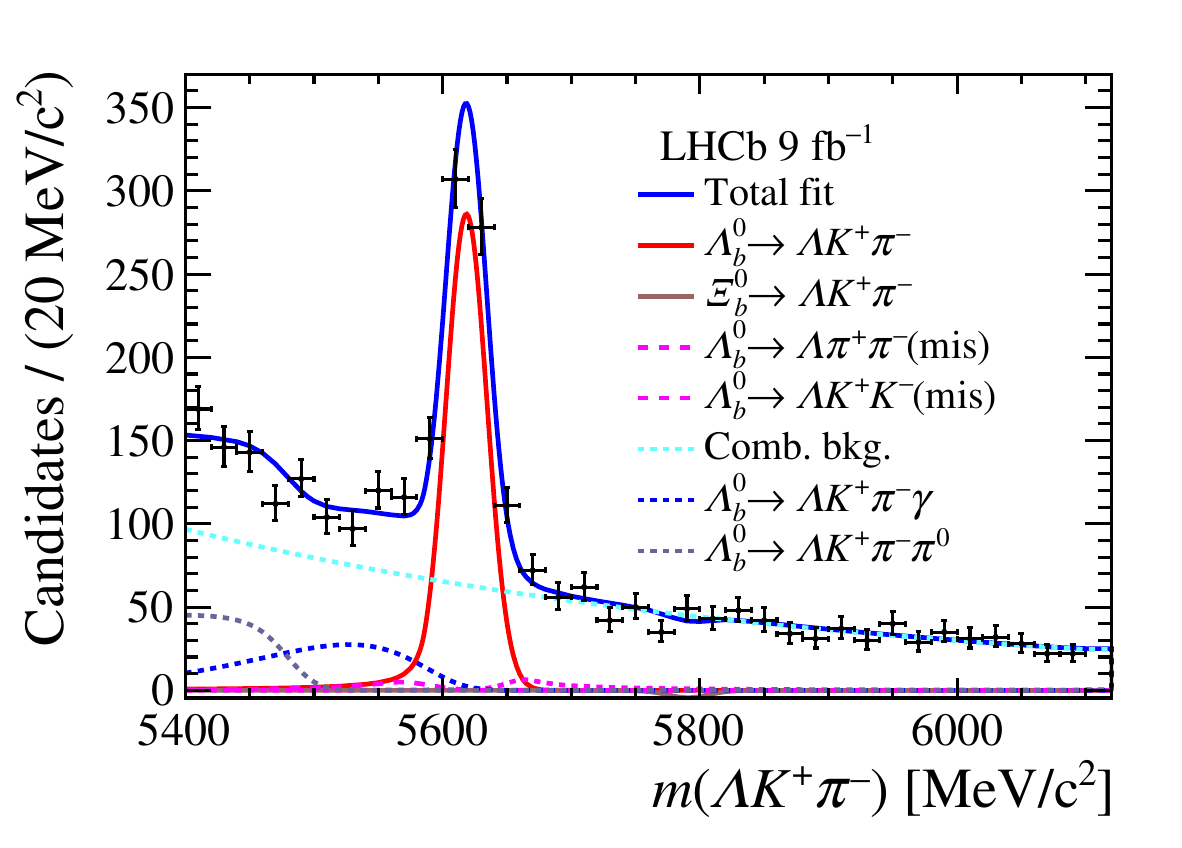}\put(-125,88){(c)}
    \includegraphics[width=0.35\columnwidth]{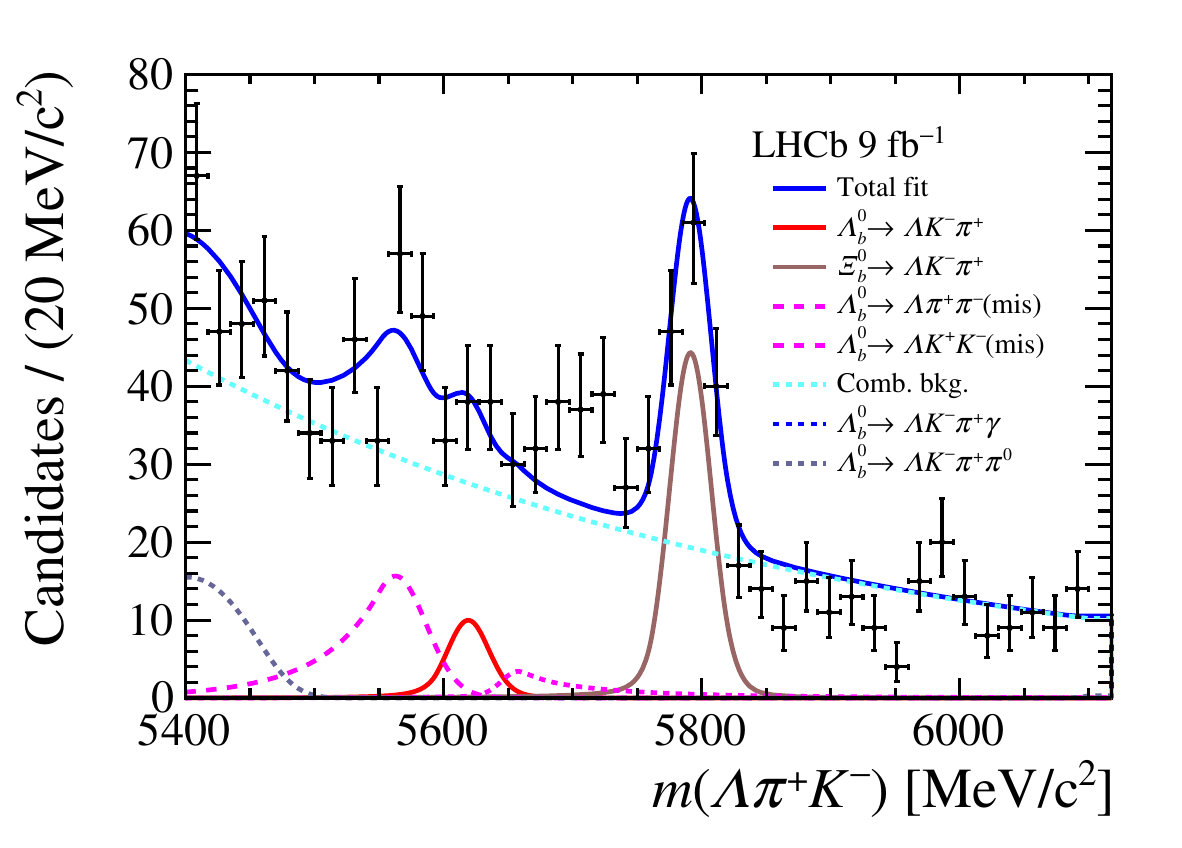}\put(-125,88){(d)}\\
    \includegraphics[width=0.35\columnwidth]{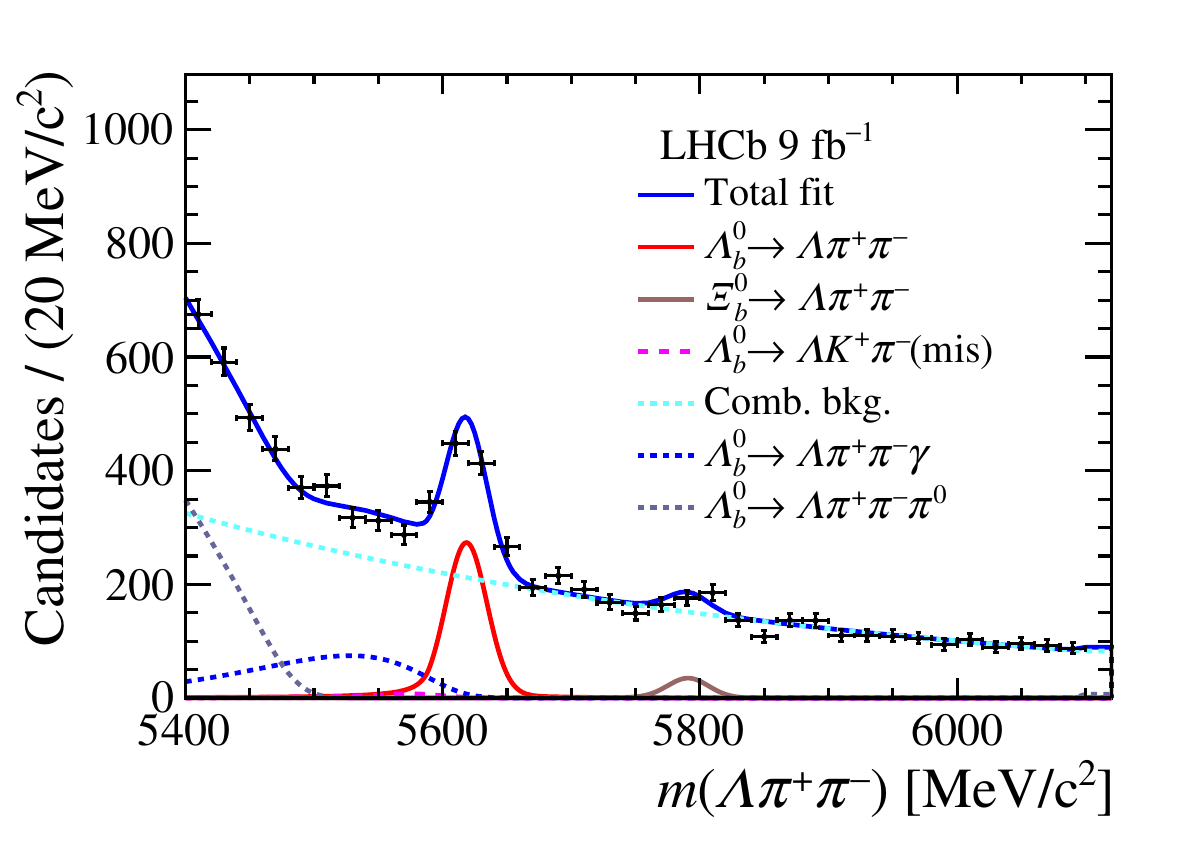}\put(-125,88){(e)}
    \includegraphics[width=0.35\columnwidth]{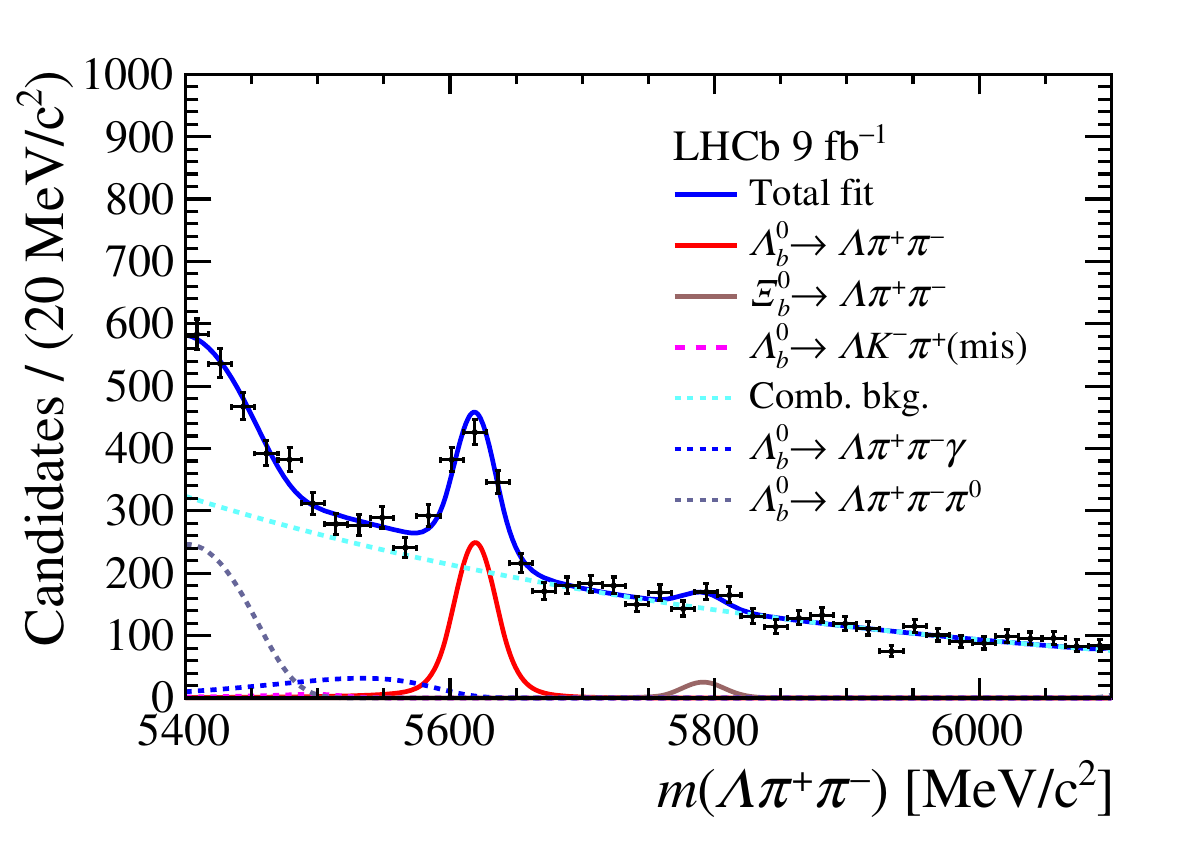}\put(-125,88){(f)}

  \end{center}
  \caption{Distributions of   $m(\Lz h^+h^-)$ for (a)(b) $\Lb(\Xibz)\to\Lz \Kp\Km$, \mbox{(c) $\decay{\Lb}{\Lz K^+\pi^-}$}, \mbox{(d) $\decay{\Xibz}{\Lz K^-\pi^+}$} and (e)(f) $\Lb(\Xibz)\to\Lz \Kp\Km$ decay modes, together with the fit results, where (a)(c)(e) are selected with the  $N_{\rm{S}}/(\sqrt{N_{\rm{B}}+N_{\rm{S}}})$ FoM, focusing on the $\Lb$ studies, while (b)(d)(f) are selected with the  $N_{\rm{S}}/(\sqrt{N_{\rm{B}}} + 2.5)$ FoM for $\Xibz$ studies.}
  \label{fig:mass spectrum}
\end{figure}

Figure~\ref{fig:mass spectrum} shows the mass spectra used to obtain yields of signal channels for branching fraction calculations of (a)(b) $\Lb(\Xibz)\to\Lz \Kp\Km$, (c) \LzKP, (d) \XiPK, and \mbox{(e)(f) $\Lb(\Xibz)\to\Lz \Kp\Km$} decay modes. The same BDT classifier is used in selecting candidates for the $\Lb$ and $\Xibz$ modes, but with a different figure-of-merit (FoM) used to choose the optimal requirement. Due to the relatively smaller number of $\Xibz$ signal yields, when determining its selection criteria the $N_{\rm{S}}/(\sqrt{N_{\rm{B}}} + 2.5)$ FoM method is applied, as shown in Fig.~\ref{fig:mass spectrum} (b)(d)(f), whereas when studying the $\Lb$ modes, the $N_{\rm{S}}/\sqrt{N_{\rm{S}} + N_{\rm{B}}}$ FoM method is applied, as shown in Fig.~\ref{fig:mass spectrum} (a)(c)(e).

\clearpage

\section{Summary tables for correction terms}

Table~\ref{tab:correction_terms} lists the production asymmetry difference $\Delta A_\text{P}$ and detection asymmetry difference $\Delta A_\text{exp}$ for each decay mode with respect to control mode.
\begin{table}[!b]
    \centering
    \caption{Production asymmetry difference $\Delta A_\text{P}$ and  detection asymmetry difference $\Delta A_\text{exp}$ for each decay mode. The uncertainties from these asymmetries are propagated into the phase-space integrated $\Delta\ACP$ as systematic uncertainties. }
    \begin{tabular}{lccccccccccc}
    \hline	Channel		&	$\Delta A_\text{P}$ [\%]			&	$\Delta A_\text{exp}$ [\%]			\\	\hline
\LzPP	&		$\phantom{-}0.1	\pm	0.1$	&	$0.1	\pm	0.9$		\\
\LzKP	&	$\phantom{-}0.2	\pm	0.2$	&	$1.4	\pm	1.0$		    \\
\LzKK	&   $-0.2	\pm	0.2$	&	$0.0	\pm	0.9$		    \\
\XiPK	&	$-5.2	\pm 	4.0$	&	$0.3	\pm 	1.6$		 \\ \hline
    \end{tabular}
    \label{tab:correction_terms}
\end{table}

\clearpage

% \input{acknowledgements}

% \clearpage

% \input{0Paper_Supplementary}
 
 \clearpage
 \ifx\mcitethebibliography\mciteundefinedmacro
\PackageError{LHCb.bst}{mciteplus.sty has not been loaded}
{This bibstyle requires the use of the mciteplus package.}\fi
\providecommand{\href}[2]{#2}

% \bibliographystyle{LHCb}
% \bibliography{main,standard,LHCb-PAPER,LHCb-CONF,LHCb-DP,LHCb-TDR}

\begin{mcitethebibliography}{10}
\mciteSetBstSublistMode{n}
\mciteSetBstMaxWidthForm{subitem}{\alph{mcitesubitemcount})}
\mciteSetBstSublistLabelBeginEnd{\mcitemaxwidthsubitemform\space}
{\relax}{\relax}

\bibitem{HFLAV21}
Heavy Flavor Averaging Group, Y.~Amhis {\em et~al.}, \ifthenelse{\boolean{articletitles}}{\emph{{Averages of $b$-hadron, $c$-hadron, and $\tau$-lepton properties as of 2021}}, }{}\href{https://doi.org/10.1103/PhysRevD.107.052008}{Phys.\ Rev.\  \textbf{D107} (2023) 052008}, \href{http://arxiv.org/abs/2206.07501}{{\normalfont\ttfamily arXiv:2206.07501}}, {updated results and plots available at \href{https://hflav.web.cern.ch}{{\texttt{https://hflav.web.cern.ch}}}}\relax
\mciteBstWouldAddEndPuncttrue
\mciteSetBstMidEndSepPunct{\mcitedefaultmidpunct}
{\mcitedefaultendpunct}{\mcitedefaultseppunct}\relax
\EndOfBibitem
\bibitem{Riotto:1999yt}
A.~Riotto and M.~Trodden, \ifthenelse{\boolean{articletitles}}{\emph{{Recent progress in baryogenesis}}, }{}\href{https://doi.org/10.1146/annurev.nucl.49.1.35}{Ann.\ Rev.\ Nucl.\ Part.\ Sci.\  \textbf{49} (1999) 35}, \href{http://arxiv.org/abs/hep-ph/9901362}{{\normalfont\ttfamily arXiv:hep-ph/9901362}}\relax
\mciteBstWouldAddEndPuncttrue
\mciteSetBstMidEndSepPunct{\mcitedefaultmidpunct}
{\mcitedefaultendpunct}{\mcitedefaultseppunct}\relax
\EndOfBibitem
\bibitem{BESIII:2022qax}
BESIII collaboration, M.~Ablikim {\em et~al.}, \ifthenelse{\boolean{articletitles}}{\emph{{Precise measurements of decay parameters and $CP$ asymmetry with entangled $\Lz-\bar{\Lz}$ pairs}}, }{}\href{https://doi.org/10.1103/PhysRevLett.129.131801}{Phys.\ Rev.\ Lett.\  \textbf{129} (2022) 131801}, \href{http://arxiv.org/abs/2204.11058}{{\normalfont\ttfamily arXiv:2204.11058}}\relax
\mciteBstWouldAddEndPuncttrue
\mciteSetBstMidEndSepPunct{\mcitedefaultmidpunct}
{\mcitedefaultendpunct}{\mcitedefaultseppunct}\relax
\EndOfBibitem
\bibitem{LHCb-PAPER-2013-061}
LHCb collaboration, R.~Aaij {\em et~al.}, \ifthenelse{\boolean{articletitles}}{\emph{{Searches for \Lb and \Xibz decays to $\KS\proton\pim$ and $\KS\proton\Km$ final states with first observation of the \mbox{\decay{\Lb}{\KS\proton\pim}} decay}}, }{}\href{https://doi.org/10.1007/JHEP04(2014)087}{JHEP \textbf{04} (2014) 087}, \href{http://arxiv.org/abs/1402.0770}{{\normalfont\ttfamily arXiv:1402.0770}}\relax
\mciteBstWouldAddEndPuncttrue
\mciteSetBstMidEndSepPunct{\mcitedefaultmidpunct}
{\mcitedefaultendpunct}{\mcitedefaultseppunct}\relax
\EndOfBibitem
\bibitem{LHCb-PAPER-2014-020}
LHCb collaboration, R.~Aaij {\em et~al.}, \ifthenelse{\boolean{articletitles}}{\emph{{Observation of the \mbox{\decay{\Lb}{\jpsi\proton\pim}} decay}}, }{}\href{https://doi.org/10.1007/JHEP07(2014)103}{JHEP \textbf{07} (2014) 103}, \href{http://arxiv.org/abs/1406.0755}{{\normalfont\ttfamily arXiv:1406.0755}}\relax
\mciteBstWouldAddEndPuncttrue
\mciteSetBstMidEndSepPunct{\mcitedefaultmidpunct}
{\mcitedefaultendpunct}{\mcitedefaultseppunct}\relax
\EndOfBibitem
\bibitem{LHCb-PAPER-2016-030}
LHCb collaboration, R.~Aaij {\em et~al.}, \ifthenelse{\boolean{articletitles}}{\emph{{Measurement of matter-antimatter differences in beauty baryon decays}}, }{}\href{https://doi.org/10.1038/nphys4021}{Nature Physics \textbf{13} (2017) 391}, \href{http://arxiv.org/abs/1609.05216}{{\normalfont\ttfamily arXiv:1609.05216}}\relax
\mciteBstWouldAddEndPuncttrue
\mciteSetBstMidEndSepPunct{\mcitedefaultmidpunct}
{\mcitedefaultendpunct}{\mcitedefaultseppunct}\relax
\EndOfBibitem
\bibitem{LHCb-PAPER-2018-001}
LHCb collaboration, R.~Aaij {\em et~al.}, \ifthenelse{\boolean{articletitles}}{\emph{{Search for \CP violation using triple product asymmetries in \mbox{\decay{\Lb}{p\Km\pip\pim}}, \mbox{\decay{\Lb}{p\Km\Kp\Km}}, and \mbox{\decay{\Xires^0_b}{p\Km\Km\pip}} decays}}, }{}\href{https://doi.org/10.1007/JHEP08(2018)039}{JHEP \textbf{08} (2018) 039}, \href{http://arxiv.org/abs/1805.03941}{{\normalfont\ttfamily arXiv:1805.03941}}\relax
\mciteBstWouldAddEndPuncttrue
\mciteSetBstMidEndSepPunct{\mcitedefaultmidpunct}
{\mcitedefaultendpunct}{\mcitedefaultseppunct}\relax
\EndOfBibitem
\bibitem{LHCb-PAPER-2018-044}
LHCb collaboration, R.~Aaij {\em et~al.}, \ifthenelse{\boolean{articletitles}}{\emph{{Measurement of \CP asymmetries in charmless four-body \Lb and \Xibz decays}}, }{}\href{https://doi.org/10.1140/epjc/s10052-019-7218-1}{Eur.\ Phys.\ J.\  \textbf{C79} (2019) 745}, \href{http://arxiv.org/abs/1903.06792}{{\normalfont\ttfamily arXiv:1903.06792}}\relax
\mciteBstWouldAddEndPuncttrue
\mciteSetBstMidEndSepPunct{\mcitedefaultmidpunct}
{\mcitedefaultendpunct}{\mcitedefaultseppunct}\relax
\EndOfBibitem
\bibitem{LHCb-PAPER-2019-028}
LHCb collaboration, R.~Aaij {\em et~al.}, \ifthenelse{\boolean{articletitles}}{\emph{{Search for \CP violation and observation of $P$ violation in \decay{\Lb}{p\pim\pip\pim} decays}}, }{}\href{https://doi.org/10.1103/PhysRevD.102.051101}{Phys.\ Rev.\  \textbf{D102} (2020) 051101}, \href{http://arxiv.org/abs/1912.10741}{{\normalfont\ttfamily arXiv:1912.10741}}\relax
\mciteBstWouldAddEndPuncttrue
\mciteSetBstMidEndSepPunct{\mcitedefaultmidpunct}
{\mcitedefaultendpunct}{\mcitedefaultseppunct}\relax
\EndOfBibitem
\bibitem{LHCb-PAPER-2021-027}
LHCb collaboration, R.~Aaij {\em et~al.}, \ifthenelse{\boolean{articletitles}}{\emph{{Observation of the suppressed $\Lb \to \D \Pp \Km$ decay with $\D \to \Kp \pim$ and measurement of its \CP asymmetry}}, }{}\href{https://doi.org/10.1103/PhysRevD.104.112008}{Phys.\ Rev.\  \textbf{D104} (2021) 112008}, \href{http://arxiv.org/abs/2109.02621}{{\normalfont\ttfamily arXiv:2109.02621}}\relax
\mciteBstWouldAddEndPuncttrue
\mciteSetBstMidEndSepPunct{\mcitedefaultmidpunct}
{\mcitedefaultendpunct}{\mcitedefaultseppunct}\relax
\EndOfBibitem
\bibitem{LHCb-PAPER-2016-004}
LHCb collaboration, R.~Aaij {\em et~al.}, \ifthenelse{\boolean{articletitles}}{\emph{{Observations of \mbox{\decay{\Lb}{\Lz\Kp\pim}} and \mbox{\decay{\Lb}{\Lz\Kp\Km}} decays and searches for other \Lb and \Xibz decays to $\Lz h^+h^-$ final states}}, }{}\href{https://doi.org/10.1007/JHEP05(2016)081}{JHEP \textbf{05} (2016) 081}, \href{http://arxiv.org/abs/1603.00413}{{\normalfont\ttfamily arXiv:1603.00413}}\relax
\mciteBstWouldAddEndPuncttrue
\mciteSetBstMidEndSepPunct{\mcitedefaultmidpunct}
{\mcitedefaultendpunct}{\mcitedefaultseppunct}\relax
\EndOfBibitem
\bibitem{LHCb-PAPER-2016-059}
LHCb collaboration, R.~Aaij {\em et~al.}, \ifthenelse{\boolean{articletitles}}{\emph{{Observation of the decay \mbox{\decay{\Lb}{\proton\Km\mumu}} and search for \CP violation}}, }{}\href{https://doi.org/10.1007/JHEP06(2017)108}{JHEP \textbf{06} (2017) 108}, \href{http://arxiv.org/abs/1703.00256}{{\normalfont\ttfamily arXiv:1703.00256}}\relax
\mciteBstWouldAddEndPuncttrue
\mciteSetBstMidEndSepPunct{\mcitedefaultmidpunct}
{\mcitedefaultendpunct}{\mcitedefaultseppunct}\relax
\EndOfBibitem
\bibitem{LHCb-PAPER-2018-025}
LHCb collaboration, R.~Aaij {\em et~al.}, \ifthenelse{\boolean{articletitles}}{\emph{{Search for \CP violation in \mbox{\decay{\Lb}{p\Km} and \mbox{\decay{\Lb}{p\pim}}} decays}}, }{}\href{https://doi.org/10.1016/j.physletb.2018.10.039}{Phys.\ Lett.\  \textbf{B784} (2018) 101}, \href{http://arxiv.org/abs/1807.06544}{{\normalfont\ttfamily arXiv:1807.06544}}\relax
\mciteBstWouldAddEndPuncttrue
\mciteSetBstMidEndSepPunct{\mcitedefaultmidpunct}
{\mcitedefaultendpunct}{\mcitedefaultseppunct}\relax
\EndOfBibitem
\bibitem{LHCb-PAPER-2020-017}
LHCb collaboration, R.~Aaij {\em et~al.}, \ifthenelse{\boolean{articletitles}}{\emph{{Search for \CP violation in $\Xibm \to p K^- K^- decays$}}, }{}\href{https://doi.org/10.1103/PhysRevD.104.052010}{Phys.\ Rev.\  \textbf{D104} (2021) 052010}, \href{http://arxiv.org/abs/2104.15074}{{\normalfont\ttfamily arXiv:2104.15074}}\relax
\mciteBstWouldAddEndPuncttrue
\mciteSetBstMidEndSepPunct{\mcitedefaultmidpunct}
{\mcitedefaultendpunct}{\mcitedefaultseppunct}\relax
\EndOfBibitem
\bibitem{LHCb-PAPER-2021-030}
LHCb collaboration, R.~Aaij {\em et~al.}, \ifthenelse{\boolean{articletitles}}{\emph{{Measurement of the photon polarization in $\Lb \to \Lz \gamma$ decays}}, }{}\href{https://doi.org/10.1103/PhysRevD.105.L051104}{Phys.\ Rev.\  \textbf{D105} (2022) 051104}, \href{http://arxiv.org/abs/2111.10194}{{\normalfont\ttfamily arXiv:2111.10194}}\relax
\mciteBstWouldAddEndPuncttrue
\mciteSetBstMidEndSepPunct{\mcitedefaultmidpunct}
{\mcitedefaultendpunct}{\mcitedefaultseppunct}\relax
\EndOfBibitem
\bibitem{LHCb-PAPER-2016-002}
LHCb collaboration, R.~Aaij {\em et~al.}, \ifthenelse{\boolean{articletitles}}{\emph{{Observation of the \mbox{\decay{\Lb}{\Lz\phi}} decay}}, }{}\href{https://doi.org/10.1016/j.physletb.2016.05.077}{Phys.\ Lett.\  \textbf{B759} (2016) 282}, \href{http://arxiv.org/abs/1603.02870}{{\normalfont\ttfamily arXiv:1603.02870}}\relax
\mciteBstWouldAddEndPuncttrue
\mciteSetBstMidEndSepPunct{\mcitedefaultmidpunct}
{\mcitedefaultendpunct}{\mcitedefaultseppunct}\relax
\EndOfBibitem
\bibitem{LHCb-PAPER-2017-044}
LHCb collaboration, R.~Aaij {\em et~al.}, \ifthenelse{\boolean{articletitles}}{\emph{{Search for \CP violation in \mbox{\decay{\Lc}{p \Km \Kp}} and \mbox{\decay{\Lc}{p\pim\pip}} decays}}, }{}\href{https://doi.org/10.1007/JHEP03(2018)182}{JHEP \textbf{03} (2018) 182}, \href{http://arxiv.org/abs/1712.07051}{{\normalfont\ttfamily arXiv:1712.07051}}\relax
\mciteBstWouldAddEndPuncttrue
\mciteSetBstMidEndSepPunct{\mcitedefaultmidpunct}
{\mcitedefaultendpunct}{\mcitedefaultseppunct}\relax
\EndOfBibitem
\bibitem{LHCb-PAPER-2019-026}
LHCb collaboration, R.~Aaij {\em et~al.}, \ifthenelse{\boolean{articletitles}}{\emph{{Search for \CP violation in \mbox{\decay{\Xicp}{p\Km\pip}} decays with model-independent techniques}}, }{}\href{https://doi.org/10.1140/epjc/s10052-020-8365-0}{Eur.\ Phys.\ J.\  \textbf{C80} (2020) 986}, \href{http://arxiv.org/abs/2006.03145}{{\normalfont\ttfamily arXiv:2006.03145}}\relax
\mciteBstWouldAddEndPuncttrue
\mciteSetBstMidEndSepPunct{\mcitedefaultmidpunct}
{\mcitedefaultendpunct}{\mcitedefaultseppunct}\relax
\EndOfBibitem
\bibitem{LHCb-PAPER-2021-049}
LHCb collaboration, R.~Aaij {\em et~al.}, \ifthenelse{\boolean{articletitles}}{\emph{{Direct \CP violation in charmless three-body decays of $B^{\pm}$ mesons}}, }{}\href{https://doi.org/10.1103/PhysRevD.108.012008}{Phys.\ Rev.\  \textbf{D108} (2023) 012008}, \href{http://arxiv.org/abs/2206.07622}{{\normalfont\ttfamily arXiv:2206.07622}}\relax
\mciteBstWouldAddEndPuncttrue
\mciteSetBstMidEndSepPunct{\mcitedefaultmidpunct}
{\mcitedefaultendpunct}{\mcitedefaultseppunct}\relax
\EndOfBibitem
\bibitem{LHCb-PAPER-2018-051}
LHCb collaboration, R.~Aaij {\em et~al.}, \ifthenelse{\boolean{articletitles}}{\emph{{Amplitude analysis of \mbox{\decay{\Bpm}{\pipm\Kp\Km}} decays}}, }{}\href{https://doi.org/10.1103/PhysRevLett.123.231802}{Phys.\ Rev.\ Lett.\  \textbf{123} (2019) 231802}, \href{http://arxiv.org/abs/1905.09244}{{\normalfont\ttfamily arXiv:1905.09244}}\relax
\mciteBstWouldAddEndPuncttrue
\mciteSetBstMidEndSepPunct{\mcitedefaultmidpunct}
{\mcitedefaultendpunct}{\mcitedefaultseppunct}\relax
\EndOfBibitem
\bibitem{LHCb-PAPER-2014-044}
LHCb collaboration, R.~Aaij {\em et~al.}, \ifthenelse{\boolean{articletitles}}{\emph{{Measurement of \CP violation in the three-body phase space of charmless \Bpm decays}}, }{}\href{https://doi.org/10.1103/PhysRevD.90.112004}{Phys.\ Rev.\  \textbf{D90} (2014) 112004}, \href{http://arxiv.org/abs/1408.5373}{{\normalfont\ttfamily arXiv:1408.5373}}\relax
\mciteBstWouldAddEndPuncttrue
\mciteSetBstMidEndSepPunct{\mcitedefaultmidpunct}
{\mcitedefaultendpunct}{\mcitedefaultseppunct}\relax
\EndOfBibitem
\bibitem{Geng:2016gul}
C.~Q. Geng, Y.~K. Hsiao, Y.-H. Lin, and Y.~Yu, \ifthenelse{\boolean{articletitles}}{\emph{{Study of $\Lb\rightarrow \Lz (\phi ,\eta ^{(\prime )})$ and $\Lb\rightarrow \Lz K^+K^-$ decays}}, }{}\href{https://doi.org/10.1140/epjc/s10052-016-4255-x}{Eur.\ Phys.\ J.\  \textbf{C76} (2016) 399}, \href{http://arxiv.org/abs/1603.06682}{{\normalfont\ttfamily arXiv:1603.06682}}\relax
\mciteBstWouldAddEndPuncttrue
\mciteSetBstMidEndSepPunct{\mcitedefaultmidpunct}
{\mcitedefaultendpunct}{\mcitedefaultseppunct}\relax
\EndOfBibitem
\bibitem{PhysRevD.69.017901}
S.~Arunagiri and C.~Q. Geng, \ifthenelse{\boolean{articletitles}}{\emph{{T violating triple product asymmetries in $\Lb \to \Lz \pip \pim$ decay}}, }{}\href{https://doi.org/10.1103/PhysRevD.69.017901}{Phys.\ Rev.\  \textbf{D69} (2004) 017901}, \href{http://arxiv.org/abs/hep-ph/0307307}{{\normalfont\ttfamily arXiv:hep-ph/0307307}}\relax
\mciteBstWouldAddEndPuncttrue
\mciteSetBstMidEndSepPunct{\mcitedefaultmidpunct}
{\mcitedefaultendpunct}{\mcitedefaultseppunct}\relax
\EndOfBibitem
\bibitem{Guo:1998eg}
X.-H. Guo and A.~W. Thomas, \ifthenelse{\boolean{articletitles}}{\emph{{Direct CP violation in $\Lb \to n_{\Lz} \pip \pim$ decays via $\rho - \omega$ mixing}}, }{}\href{https://doi.org/10.1103/PhysRevD.58.096013}{Phys.\ Rev.\  \textbf{D58} (1998) 096013}, \href{http://arxiv.org/abs/hep-ph/9805332}{{\normalfont\ttfamily arXiv:hep-ph/9805332}}\relax
\mciteBstWouldAddEndPuncttrue
\mciteSetBstMidEndSepPunct{\mcitedefaultmidpunct}
{\mcitedefaultendpunct}{\mcitedefaultseppunct}\relax
\EndOfBibitem
\bibitem{PhysRevD.99.054020}
J.~Zhu, Z.-T. Wei, and H.-W. Ke, \ifthenelse{\boolean{articletitles}}{\emph{{Semileptonic and nonleptonic weak decays of $\Lb$}}, }{}\href{https://doi.org/10.1103/PhysRevD.99.054020}{Phys.\ Rev.\  \textbf{D99} (2019) 054020}\relax
\mciteBstWouldAddEndPuncttrue
\mciteSetBstMidEndSepPunct{\mcitedefaultmidpunct}
{\mcitedefaultendpunct}{\mcitedefaultseppunct}\relax
\EndOfBibitem
\bibitem{Rui:2022jff}
Z.~Rui, J.-M. Li, and C.-Q. Zhang, \ifthenelse{\boolean{articletitles}}{\emph{{Estimates of exchange topological contributions and CP-violating observables in $\Lb\to\Lz\phi$ decay}}, }{}\href{https://doi.org/10.1103/PhysRevD.107.053009}{Phys.\ Rev.\  \textbf{D107} (2023) 053009}, \href{http://arxiv.org/abs/2210.15357}{{\normalfont\ttfamily arXiv:2210.15357}}\relax
\mciteBstWouldAddEndPuncttrue
\mciteSetBstMidEndSepPunct{\mcitedefaultmidpunct}
{\mcitedefaultendpunct}{\mcitedefaultseppunct}\relax
\EndOfBibitem
\bibitem{PhysRevD.95.093001}
Y.~K. Hsiao, Y.~Yao, and C.~Q. Geng, \ifthenelse{\boolean{articletitles}}{\emph{{Charmless two-body antitriplet $b$-baryon decays}}, }{}\href{https://doi.org/10.1103/PhysRevD.95.093001}{Phys.\ Rev.\  \textbf{D95} (2017) 093001}\relax
\mciteBstWouldAddEndPuncttrue
\mciteSetBstMidEndSepPunct{\mcitedefaultmidpunct}
{\mcitedefaultendpunct}{\mcitedefaultseppunct}\relax
\EndOfBibitem
\bibitem{Wang:2024}
J.-P. Wang and F.-S. Yu, \ifthenelse{\boolean{articletitles}}{\emph{{ $\CP$ violation of baryon decays with $N\pi$ rescatterings}}, }{}\href{https://doi.org/10.1088/1674-1137/ad75f4}{Chinese Physics \textbf{C48} (2024) 101001}\relax
\mciteBstWouldAddEndPuncttrue
\mciteSetBstMidEndSepPunct{\mcitedefaultmidpunct}
{\mcitedefaultendpunct}{\mcitedefaultseppunct}\relax
\EndOfBibitem
\bibitem{LHCb-DP-2008-001}
LHCb collaboration, A.~A. Alves~Jr.\ {\em et~al.}, \ifthenelse{\boolean{articletitles}}{\emph{{The \lhcb detector at the LHC}}, }{}\href{https://doi.org/10.1088/1748-0221/3/08/S08005}{JINST \textbf{3} (2008) S08005}\relax
\mciteBstWouldAddEndPuncttrue
\mciteSetBstMidEndSepPunct{\mcitedefaultmidpunct}
{\mcitedefaultendpunct}{\mcitedefaultseppunct}\relax
\EndOfBibitem
\bibitem{LHCb-DP-2014-002}
LHCb collaboration, R.~Aaij {\em et~al.}, \ifthenelse{\boolean{articletitles}}{\emph{{LHCb detector performance}}, }{}\href{https://doi.org/10.1142/S0217751X15300227}{Int.\ J.\ Mod.\ Phys.\  \textbf{A30} (2015) 1530022}, \href{http://arxiv.org/abs/1412.6352}{{\normalfont\ttfamily arXiv:1412.6352}}\relax
\mciteBstWouldAddEndPuncttrue
\mciteSetBstMidEndSepPunct{\mcitedefaultmidpunct}
{\mcitedefaultendpunct}{\mcitedefaultseppunct}\relax
\EndOfBibitem
\bibitem{LHCb-DP-2013-003}
R.~Arink {\em et~al.}, \ifthenelse{\boolean{articletitles}}{\emph{{Performance of the LHCb Outer Tracker}}, }{}\href{https://doi.org/10.1088/1748-0221/9/01/P01002}{JINST \textbf{9} (2014) P01002}, \href{http://arxiv.org/abs/1311.3893}{{\normalfont\ttfamily arXiv:1311.3893}}\relax
\mciteBstWouldAddEndPuncttrue
\mciteSetBstMidEndSepPunct{\mcitedefaultmidpunct}
{\mcitedefaultendpunct}{\mcitedefaultseppunct}\relax
\EndOfBibitem
\bibitem{LHCb-DP-2017-001}
P.~d'Argent {\em et~al.}, \ifthenelse{\boolean{articletitles}}{\emph{{Improved performance of the LHCb Outer Tracker in LHC Run 2}}, }{}\href{https://doi.org/10.1088/1748-0221/12/11/P11016}{JINST \textbf{12} (2017) P11016}, \href{http://arxiv.org/abs/1708.00819}{{\normalfont\ttfamily arXiv:1708.00819}}\relax
\mciteBstWouldAddEndPuncttrue
\mciteSetBstMidEndSepPunct{\mcitedefaultmidpunct}
{\mcitedefaultendpunct}{\mcitedefaultseppunct}\relax
\EndOfBibitem
\bibitem{LHCb-DP-2012-003}
M.~Adinolfi {\em et~al.}, \ifthenelse{\boolean{articletitles}}{\emph{{Performance of the \lhcb RICH detector at the LHC}}, }{}\href{https://doi.org/10.1140/epjc/s10052-013-2431-9}{Eur.\ Phys.\ J.\  \textbf{C73} (2013) 2431}, \href{http://arxiv.org/abs/1211.6759}{{\normalfont\ttfamily arXiv:1211.6759}}\relax
\mciteBstWouldAddEndPuncttrue
\mciteSetBstMidEndSepPunct{\mcitedefaultmidpunct}
{\mcitedefaultendpunct}{\mcitedefaultseppunct}\relax
\EndOfBibitem
\bibitem{LHCb-PUB-2016-021}
L.~Anderlini {\em et~al.}, \ifthenelse{\boolean{articletitles}}{\emph{{The PIDCalib package}}, }{} \href{http://cdsweb.cern.ch/search?p=LHCb-PUB-2016-021&f=reportnumber&action_search=Search&c=LHCb+Notes} {LHCb-PUB-2016-021}, 2016\relax
\mciteBstWouldAddEndPuncttrue
\mciteSetBstMidEndSepPunct{\mcitedefaultmidpunct}
{\mcitedefaultendpunct}{\mcitedefaultseppunct}\relax
\EndOfBibitem
\bibitem{LHCb-DP-2019-001}
R.~Aaij {\em et~al.}, \ifthenelse{\boolean{articletitles}}{\emph{{Design and performance of the LHCb trigger and full real-time reconstruction in Run 2 of the LHC}}, }{}\href{https://doi.org/10.1088/1748-0221/14/04/P04013}{JINST \textbf{14} (2019) P04013}, \href{http://arxiv.org/abs/1812.10790}{{\normalfont\ttfamily arXiv:1812.10790}}\relax
\mciteBstWouldAddEndPuncttrue
\mciteSetBstMidEndSepPunct{\mcitedefaultmidpunct}
{\mcitedefaultendpunct}{\mcitedefaultseppunct}\relax
\EndOfBibitem
\bibitem{LHCb-DP-2012-004}
R.~Aaij {\em et~al.}, \ifthenelse{\boolean{articletitles}}{\emph{{The \lhcb trigger and its performance in 2011}}, }{}\href{https://doi.org/10.1088/1748-0221/8/04/P04022}{JINST \textbf{8} (2013) P04022}, \href{http://arxiv.org/abs/1211.3055}{{\normalfont\ttfamily arXiv:1211.3055}}\relax
\mciteBstWouldAddEndPuncttrue
\mciteSetBstMidEndSepPunct{\mcitedefaultmidpunct}
{\mcitedefaultendpunct}{\mcitedefaultseppunct}\relax
\EndOfBibitem
\bibitem{Sjostrand:2007gs}
T.~Sj\"{o}strand, S.~Mrenna, and P.~Skands, \ifthenelse{\boolean{articletitles}}{\emph{{A brief introduction to PYTHIA 8.1}}, }{}\href{https://doi.org/10.1016/j.cpc.2008.01.036}{Comput.\ Phys.\ Commun.\  \textbf{178} (2008) 852}, \href{http://arxiv.org/abs/0710.3820}{{\normalfont\ttfamily arXiv:0710.3820}}\relax
\mciteBstWouldAddEndPuncttrue
\mciteSetBstMidEndSepPunct{\mcitedefaultmidpunct}
{\mcitedefaultendpunct}{\mcitedefaultseppunct}\relax
\EndOfBibitem
\bibitem{Sjostrand:2006za}
T.~Sj\"{o}strand, S.~Mrenna, and P.~Skands, \ifthenelse{\boolean{articletitles}}{\emph{{PYTHIA 6.4 physics and manual}}, }{}\href{https://doi.org/10.1088/1126-6708/2006/05/026}{JHEP \textbf{05} (2006) 026}, \href{http://arxiv.org/abs/hep-ph/0603175}{{\normalfont\ttfamily arXiv:hep-ph/0603175}}\relax
\mciteBstWouldAddEndPuncttrue
\mciteSetBstMidEndSepPunct{\mcitedefaultmidpunct}
{\mcitedefaultendpunct}{\mcitedefaultseppunct}\relax
\EndOfBibitem
\bibitem{Lange:2001uf}
D.~J. Lange, \ifthenelse{\boolean{articletitles}}{\emph{{The EvtGen particle decay simulation package}}, }{}\href{https://doi.org/10.1016/S0168-9002(01)00089-4}{Nucl.\ Instrum.\ Meth.\  \textbf{A462} (2001) 152}\relax
\mciteBstWouldAddEndPuncttrue
\mciteSetBstMidEndSepPunct{\mcitedefaultmidpunct}
{\mcitedefaultendpunct}{\mcitedefaultseppunct}\relax
\EndOfBibitem
\bibitem{Golonka:2005pn}
P.~Golonka and Z.~Was, \ifthenelse{\boolean{articletitles}}{\emph{{PHOTOS Monte Carlo: A precision tool for QED corrections in $Z$ and $W$ decays}}, }{}\href{https://doi.org/10.1140/epjc/s2005-02396-4}{Eur.\ Phys.\ J.\  \textbf{C45} (2006) 97}, \href{http://arxiv.org/abs/hep-ph/0506026}{{\normalfont\ttfamily arXiv:hep-ph/0506026}}\relax
\mciteBstWouldAddEndPuncttrue
\mciteSetBstMidEndSepPunct{\mcitedefaultmidpunct}
{\mcitedefaultendpunct}{\mcitedefaultseppunct}\relax
\EndOfBibitem
\bibitem{Allison:2006ve}
Geant4 collaboration, J.~Allison {\em et~al.}, \ifthenelse{\boolean{articletitles}}{\emph{{Geant4 developments and applications}}, }{}\href{https://doi.org/10.1109/TNS.2006.869826}{IEEE Trans.\ Nucl.\ Sci.\  \textbf{53} (2006) 270}\relax
\mciteBstWouldAddEndPuncttrue
\mciteSetBstMidEndSepPunct{\mcitedefaultmidpunct}
{\mcitedefaultendpunct}{\mcitedefaultseppunct}\relax
\EndOfBibitem
\bibitem{Agostinelli:2002hh}
Geant4 collaboration, S.~Agostinelli {\em et~al.}, \ifthenelse{\boolean{articletitles}}{\emph{{Geant4: A simulation toolkit}}, }{}\href{https://doi.org/10.1016/S0168-9002(03)01368-8}{Nucl.\ Instrum.\ Meth.\  \textbf{A506} (2003) 250}\relax
\mciteBstWouldAddEndPuncttrue
\mciteSetBstMidEndSepPunct{\mcitedefaultmidpunct}
{\mcitedefaultendpunct}{\mcitedefaultseppunct}\relax
\EndOfBibitem
\bibitem{LHCb-PROC-2011-006}
M.~Clemencic {\em et~al.}, \ifthenelse{\boolean{articletitles}}{\emph{{The \lhcb simulation application, Gauss: Design, evolution and experience}}, }{}\href{https://doi.org/10.1088/1742-6596/331/3/032023}{J.\ Phys.\ Conf.\ Ser.\  \textbf{331} (2011) 032023}\relax
\mciteBstWouldAddEndPuncttrue
\mciteSetBstMidEndSepPunct{\mcitedefaultmidpunct}
{\mcitedefaultendpunct}{\mcitedefaultseppunct}\relax
\EndOfBibitem
\bibitem{Breiman}
L.~Breiman, J.~H. Friedman, R.~A. Olshen, and C.~J. Stone, {\em Classification and regression trees}, Wadsworth international group, Belmont, California, USA, 1984\relax
\mciteBstWouldAddEndPuncttrue
\mciteSetBstMidEndSepPunct{\mcitedefaultmidpunct}
{\mcitedefaultendpunct}{\mcitedefaultseppunct}\relax
\EndOfBibitem
\bibitem{Hocker:2007ht}
H.~Voss, A.~Hoecker, J.~Stelzer, and F.~Tegenfeldt, \ifthenelse{\boolean{articletitles}}{\emph{{TMVA - Toolkit for Multivariate Data Analysis with ROOT}}, }{}\href{https://doi.org/10.22323/1.050.0040}{PoS \textbf{ACAT} (2007) 040}\relax
\mciteBstWouldAddEndPuncttrue
\mciteSetBstMidEndSepPunct{\mcitedefaultmidpunct}
{\mcitedefaultendpunct}{\mcitedefaultseppunct}\relax
\EndOfBibitem
\bibitem{PDG2024}
Particle Data Group, S.~Navas {\em et~al.}, \ifthenelse{\boolean{articletitles}}{\emph{{\href{http://pdg.lbl.gov/}{Review of particle physics}}}, }{}\href{https://doi.org/10.1103/PhysRevD.110.030001}{Phys.\ Rev.\  \textbf{D110} (2024) 030001}\relax
\mciteBstWouldAddEndPuncttrue
\mciteSetBstMidEndSepPunct{\mcitedefaultmidpunct}
{\mcitedefaultendpunct}{\mcitedefaultseppunct}\relax
\EndOfBibitem
\bibitem{Skwarnicki:1986xj}
T.~Skwarnicki, {\em {A study of the radiative cascade transitions between the Upsilon-prime and Upsilon resonances}}, PhD thesis, Institute of Nuclear Physics, Krakow, 1986, {\href{http://inspirehep.net/record/230779/}{DESY-F31-86-02}}\relax
\mciteBstWouldAddEndPuncttrue
\mciteSetBstMidEndSepPunct{\mcitedefaultmidpunct}
{\mcitedefaultendpunct}{\mcitedefaultseppunct}\relax
\EndOfBibitem
\bibitem{ARGUS:1990hfq}
ARGUS collaboration, H.~Albrecht {\em et~al.}, \ifthenelse{\boolean{articletitles}}{\emph{{Search for hadronic $b \to u$ decays}}, }{}\href{https://doi.org/10.1016/0370-2693(90)91293-K}{Phys.\ Lett.\  \textbf{B241} (1990) 278}\relax
\mciteBstWouldAddEndPuncttrue
\mciteSetBstMidEndSepPunct{\mcitedefaultmidpunct}
{\mcitedefaultendpunct}{\mcitedefaultseppunct}\relax
\EndOfBibitem
\bibitem{Wilks:1938dza}
S.~S. Wilks, \ifthenelse{\boolean{articletitles}}{\emph{{The large-sample distribution of the likelihood ratio for testing composite hypotheses}}, }{}\href{https://doi.org/10.1214/aoms/1177732360}{Ann.\ Math.\ Stat.\  \textbf{9} (1938) 60}\relax
\mciteBstWouldAddEndPuncttrue
\mciteSetBstMidEndSepPunct{\mcitedefaultmidpunct}
{\mcitedefaultendpunct}{\mcitedefaultseppunct}\relax
\EndOfBibitem
\bibitem{LHCb-PAPER-2014-004}
LHCb collaboration, R.~Aaij {\em et~al.}, \ifthenelse{\boolean{articletitles}}{\emph{{Study of the kinematic dependences of \Lb production in \proton\proton collisions and a measurement of the \mbox{\decay{\Lb}{\Lc\pim}} branching fraction}}, }{}\href{https://doi.org/10.1007/JHEP08(2014)143}{JHEP \textbf{08} (2014) 143}, \href{http://arxiv.org/abs/1405.6842}{{\normalfont\ttfamily arXiv:1405.6842}}\relax
\mciteBstWouldAddEndPuncttrue
\mciteSetBstMidEndSepPunct{\mcitedefaultmidpunct}
{\mcitedefaultendpunct}{\mcitedefaultseppunct}\relax
\EndOfBibitem
\bibitem{LHCb-PAPER-2018-047}
LHCb collaboration, R.~Aaij {\em et~al.}, \ifthenelse{\boolean{articletitles}}{\emph{{Measurement of the mass and production rate of \Xibm baryons}}, }{}\href{https://doi.org/10.1103/PhysRevD.99.052006}{Phys.\ Rev.\  \textbf{D99} (2019) 052006}, \href{http://arxiv.org/abs/1901.07075}{{\normalfont\ttfamily arXiv:1901.07075}}\relax
\mciteBstWouldAddEndPuncttrue
\mciteSetBstMidEndSepPunct{\mcitedefaultmidpunct}
{\mcitedefaultendpunct}{\mcitedefaultseppunct}\relax
\EndOfBibitem
\bibitem{LHCb-PAPER-2015-037}
LHCb collaboration, R.~Aaij {\em et~al.}, \ifthenelse{\boolean{articletitles}}{\emph{{Measurement of forward \jpsi production cross-sections in \proton\proton collisions at \mbox{$\sqs=$13~\tev}}}, }{}\href{https://doi.org/10.1007/JHEP10(2015)172}{JHEP \textbf{10} (2015) 172}, Erratum \href{https://doi.org/10.1007/JHEP05(2017)063}{ibid.\   \textbf{05} (2017) 063}, \href{http://arxiv.org/abs/1509.00771}{{\normalfont\ttfamily arXiv:1509.00771}}\relax
\mciteBstWouldAddEndPuncttrue
\mciteSetBstMidEndSepPunct{\mcitedefaultmidpunct}
{\mcitedefaultendpunct}{\mcitedefaultseppunct}\relax
\EndOfBibitem
\bibitem{BESIII:2015bjk}
BESIII collaboration, M.~Ablikim {\em et~al.}, \ifthenelse{\boolean{articletitles}}{\emph{{Measurements of absolute hadronic branching fractions of $\Lc$ baryon}}, }{}\href{https://doi.org/10.1103/PhysRevLett.116.052001}{Phys.\ Rev.\ Lett.\  \textbf{116} (2016) 052001}, \href{http://arxiv.org/abs/1511.08380}{{\normalfont\ttfamily arXiv:1511.08380}}\relax
\mciteBstWouldAddEndPuncttrue
\mciteSetBstMidEndSepPunct{\mcitedefaultmidpunct}
{\mcitedefaultendpunct}{\mcitedefaultseppunct}\relax
\EndOfBibitem
\bibitem{LHCb-PAPER-2022-004}
LHCb collaboration, R.~Aaij {\em et~al.}, \ifthenelse{\boolean{articletitles}}{\emph{{Search for the rare hadronic decay $\Bs \to \proton\antiproton$}}, }{}\href{https://doi.org/10.1103/PhysRevD.108.012007}{Phys.\ Rev.\  \textbf{D108} (2023) 012007}, \href{http://arxiv.org/abs/2206.06673}{{\normalfont\ttfamily arXiv:2206.06673}}\relax
\mciteBstWouldAddEndPuncttrue
\mciteSetBstMidEndSepPunct{\mcitedefaultmidpunct}
{\mcitedefaultendpunct}{\mcitedefaultseppunct}\relax
\EndOfBibitem
\bibitem{LHCb-PAPER-2014-013}
LHCb collaboration, R.~Aaij {\em et~al.}, \ifthenelse{\boolean{articletitles}}{\emph{{Measurement of \CP asymmetry in \mbox{\decay{\Dz}{\Km\Kp}} and \mbox{\decay{\Dz}{\pim\pip}} decays}}, }{}\href{https://doi.org/10.1007/JHEP07(2014)041}{JHEP \textbf{07} (2014) 041}, \href{http://arxiv.org/abs/1405.2797}{{\normalfont\ttfamily arXiv:1405.2797}}\relax
\mciteBstWouldAddEndPuncttrue
\mciteSetBstMidEndSepPunct{\mcitedefaultmidpunct}
{\mcitedefaultendpunct}{\mcitedefaultseppunct}\relax
\EndOfBibitem
\bibitem{LHCb-PAPER-2023-018}
LHCb collaboration, R.~Aaij {\em et~al.}, \ifthenelse{\boolean{articletitles}}{\emph{{Search for prompt production of pentaquarks in open charm hadron final states}}, }{}\href{https://doi.org/10.1103/PhysRevD.110.032001}{{Phys.\ Rev.\ } \textbf{D110} (2024) 032001}, \href{http://arxiv.org/abs/2404.07131}{{\normalfont\ttfamily arXiv:2404.07131}}\relax
\mciteBstWouldAddEndPuncttrue
\mciteSetBstMidEndSepPunct{\mcitedefaultmidpunct}
{\mcitedefaultendpunct}{\mcitedefaultseppunct}\relax
\EndOfBibitem
\bibitem{Pivk:2004ty}
M.~Pivk and F.~R. Le~Diberder, \ifthenelse{\boolean{articletitles}}{\emph{{sPlot: A statistical tool to unfold data distributions}}, }{}\href{https://doi.org/10.1016/j.nima.2005.08.106}{Nucl.\ Instrum.\ Meth.\  \textbf{A555} (2005) 356}, \href{http://arxiv.org/abs/physics/0402083}{{\normalfont\ttfamily arXiv:physics/0402083}}\relax
\mciteBstWouldAddEndPuncttrue
\mciteSetBstMidEndSepPunct{\mcitedefaultmidpunct}
{\mcitedefaultendpunct}{\mcitedefaultseppunct}\relax
\EndOfBibitem
\bibitem{Santos:2013gra}
D.~Mart{\'\i}nez~Santos and F.~Dupertuis, \ifthenelse{\boolean{articletitles}}{\emph{{Mass distributions marginalized over per-event errors}}, }{}\href{https://doi.org/10.1016/j.nima.2014.06.081}{Nucl.\ Instrum.\ Meth.\  \textbf{A764} (2014) 150}, \href{http://arxiv.org/abs/1312.5000}{{\normalfont\ttfamily arXiv:1312.5000}}\relax
\mciteBstWouldAddEndPuncttrue
\mciteSetBstMidEndSepPunct{\mcitedefaultmidpunct}
{\mcitedefaultendpunct}{\mcitedefaultseppunct}\relax
\EndOfBibitem
\end{mcitethebibliography}

\newpage
% LHCb collaboration author list
% Data extracted on September 12th, 2024 at 10:52am for paper reference LHCb-PAPER-2024-043
\centerline
{\large\bf LHCb collaboration}
\begin
{flushleft}
\small
R.~Aaij$^{38}$\lhcborcid{0000-0003-0533-1952},
A.S.W.~Abdelmotteleb$^{57}$\lhcborcid{0000-0001-7905-0542},
C.~Abellan~Beteta$^{51}$,
F.~Abudin{\'e}n$^{57}$\lhcborcid{0000-0002-6737-3528},
T.~Ackernley$^{61}$\lhcborcid{0000-0002-5951-3498},
A. A. ~Adefisoye$^{69}$\lhcborcid{0000-0003-2448-1550},
B.~Adeva$^{47}$\lhcborcid{0000-0001-9756-3712},
M.~Adinolfi$^{55}$\lhcborcid{0000-0002-1326-1264},
P.~Adlarson$^{82}$\lhcborcid{0000-0001-6280-3851},
C.~Agapopoulou$^{14}$\lhcborcid{0000-0002-2368-0147}
% ...

{\footnotesize \it

% ...
$^{14}$Universit{\'e} Paris-Saclay, CNRS/IN2P3, IJCLab, Orsay, France\\
% ...
$^{38}$Nikhef National Institute for Subatomic Physics, Amsterdam, Netherlands\\
% ...
$^{47}$Instituto Galego de F{\'\i}sica de Altas Enerx{\'\i}as (IGFAE), Universidade de Santiago de Compostela, Santiago de Compostela, Spain\\
% ...
$^{51}$Physik-Institut, Universit{\"a}t Z{\"u}rich, Z{\"u}rich, Switzerland\\
% ...
$^{55}$H.H. Wills Physics Laboratory, University of Bristol, Bristol, United Kingdom\\
% ...
$^{57}$Department of Physics, University of Warwick, Coventry, United Kingdom\\
% ...
$^{60}$School of Physics and Astronomy, University of Glasgow, Glasgow, United Kingdom\\
$^{61}$Oliver Lodge Laboratory, University of Liverpool, Liverpool, United Kingdom\\
% ...
$^{69}$Syracuse University, Syracuse, NY, United States\\
% ...
$^{82}$Department of Physics and Astronomy, Uppsala University, Uppsala, Sweden, associated to $^{60}$\\
% ...
}
\end{flushleft}
\newpage
\renewcommand\thefigure{S\arabic{figure}}
\setcounter{figure}{0}
\renewcommand\thetable{S\arabic{table}} 
\setcounter{table}{0}
\section*{Supplementary Material}

\begin{figure}[h]
    \centering
    \begin{minipage}[h]{0.44\columnwidth}
        \centering
        \includegraphics[width=\columnwidth]{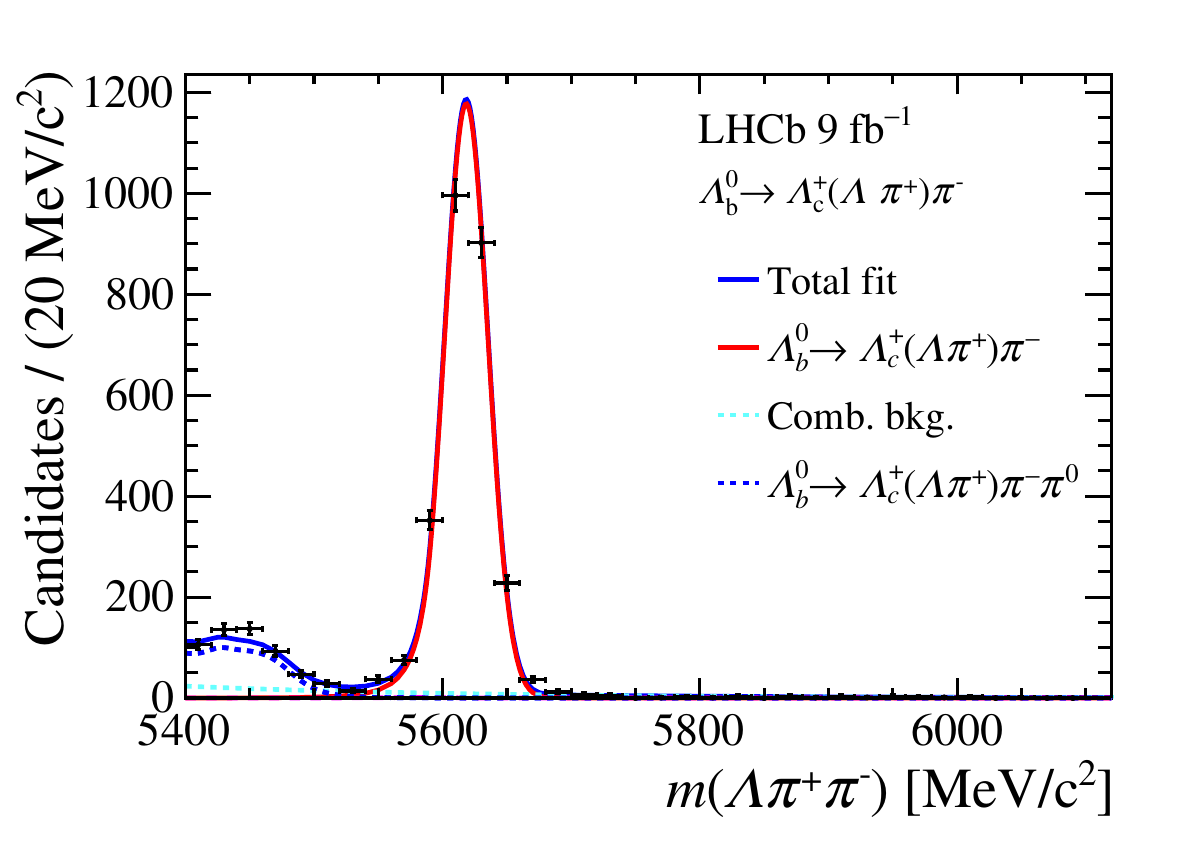}
        \put(-147,108){(a)}
    \end{minipage}
    \begin{minipage}[h]{0.44\columnwidth}
        \centering
        \includegraphics[width=\columnwidth]{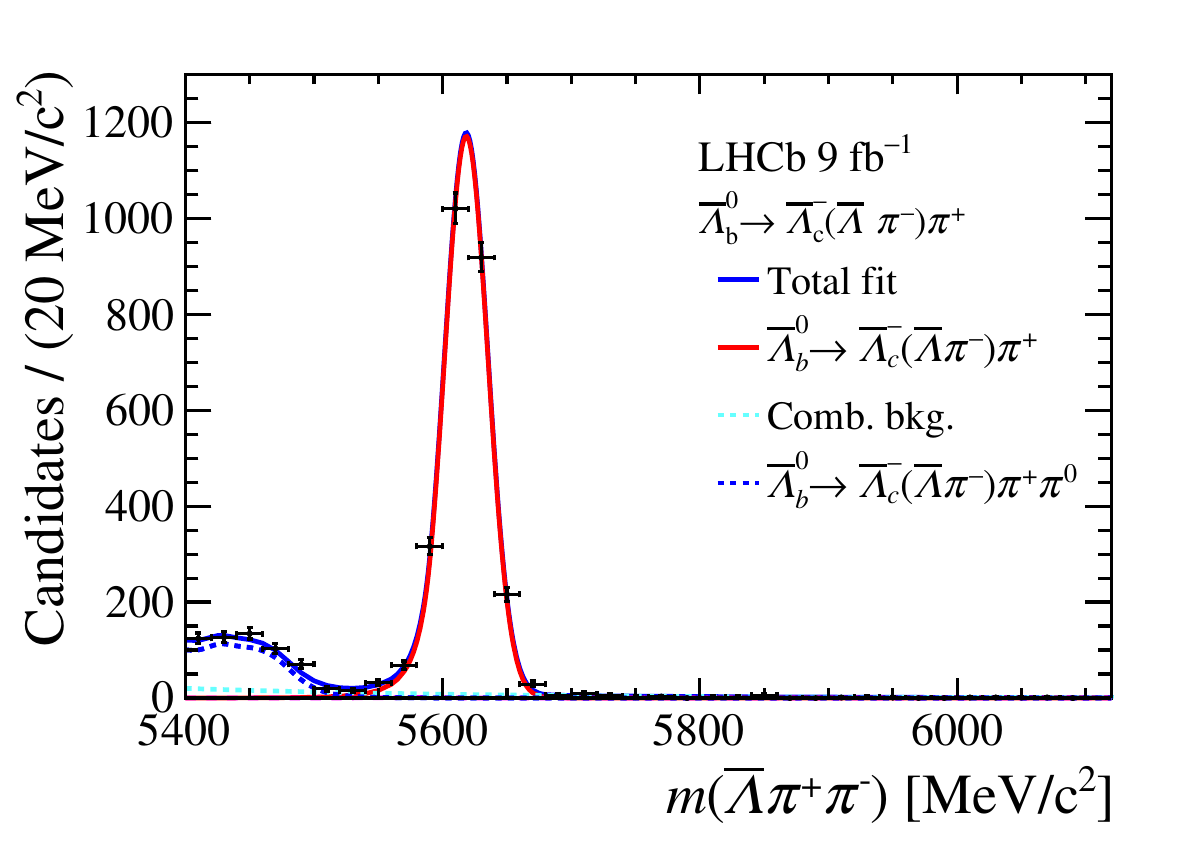}
        \put(-147,108){(b)}\\
    \end{minipage}
    
\caption{Mass distribution for the (left) $\Lb\to\Lc(\to\Lz \pip)\pim$ (right) $\Lbbar\to\Lcbar(\to\Lbar \pim)\pip$ control mode. }\label{fig:mass_ref}
\end{figure}

\begin{figure}[!b]
  \begin{center}
    \includegraphics[width=0.44\columnwidth]{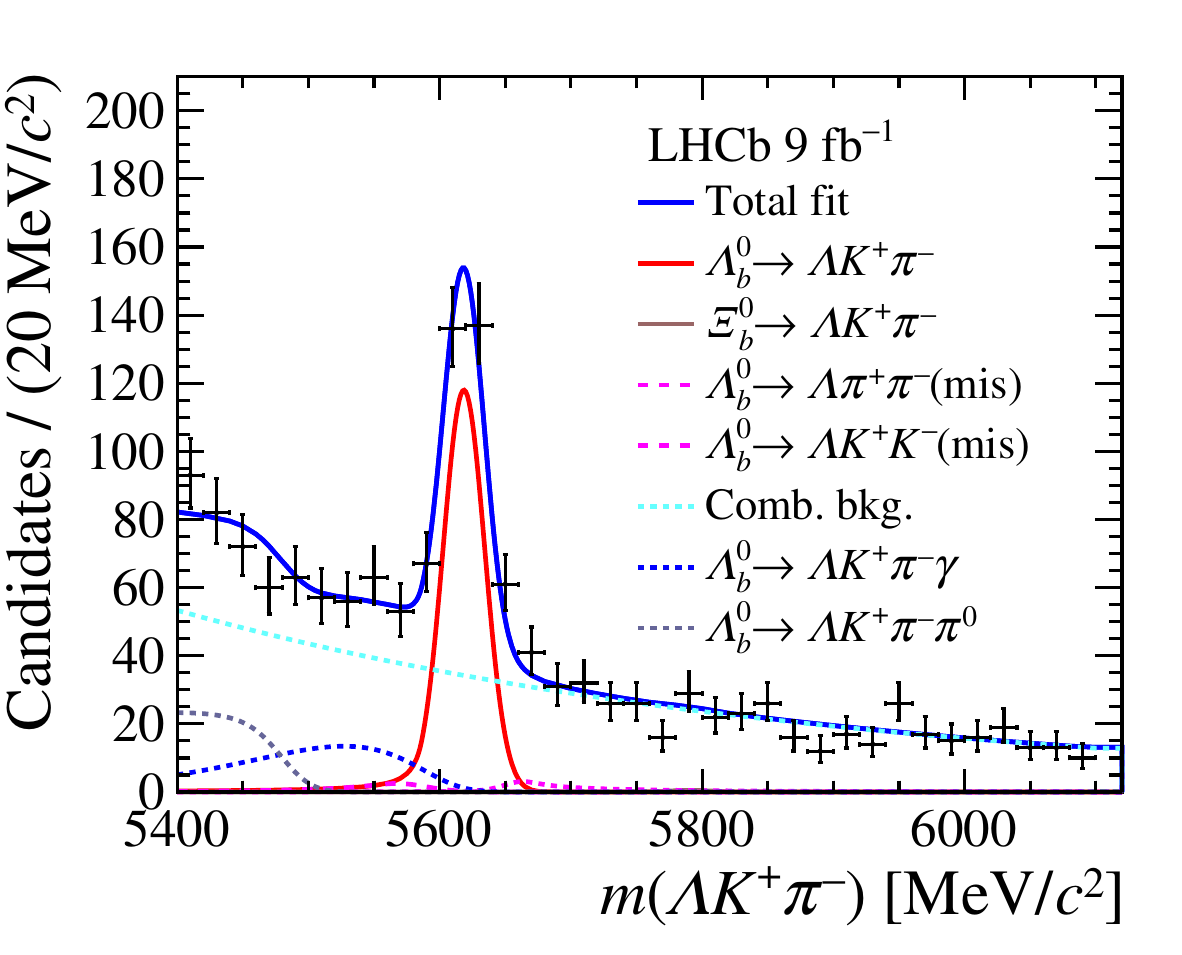}\put(-145,128){(a)}
    \includegraphics[width=0.44\columnwidth]{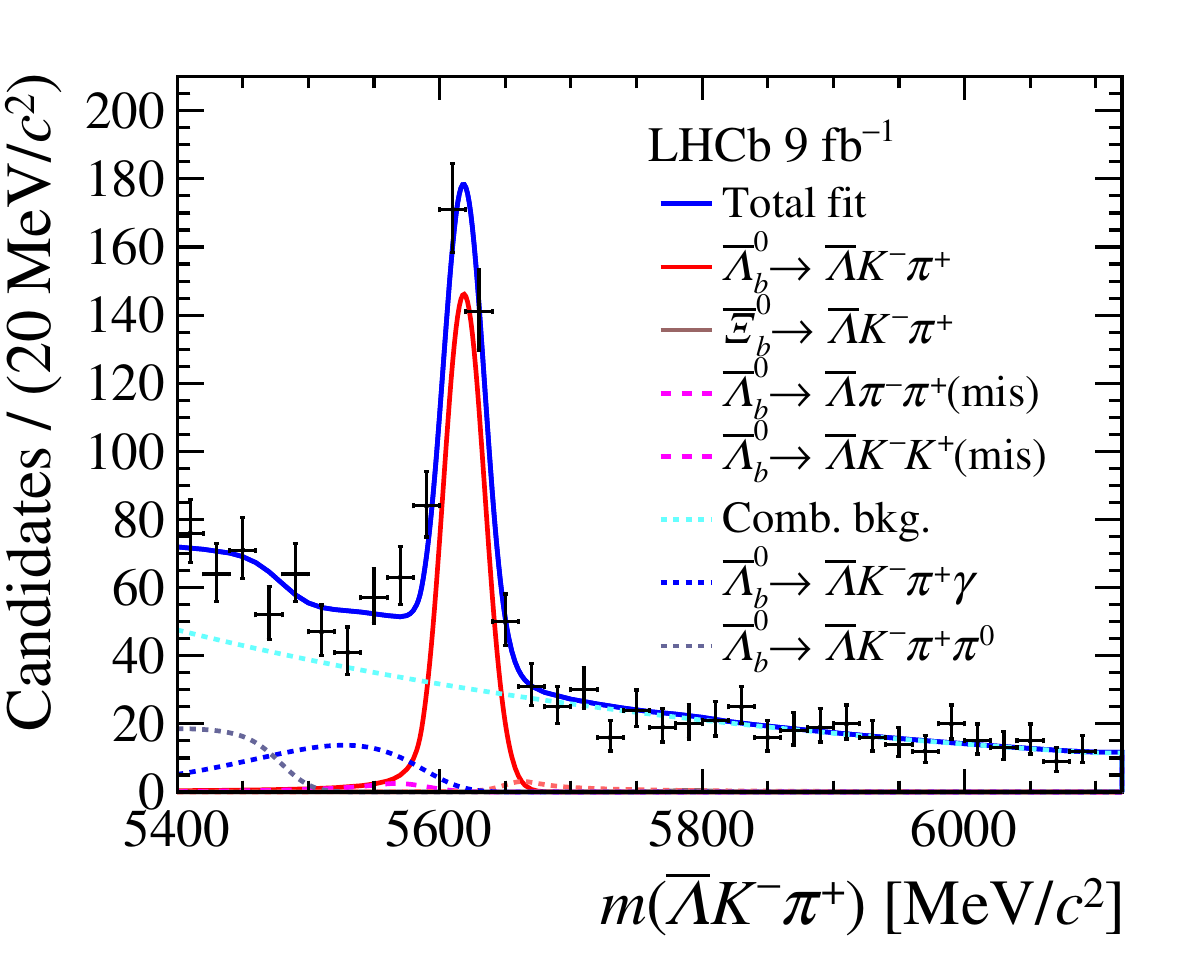}\put(-145,128){(b)}\\
    \includegraphics[width=0.44\columnwidth]{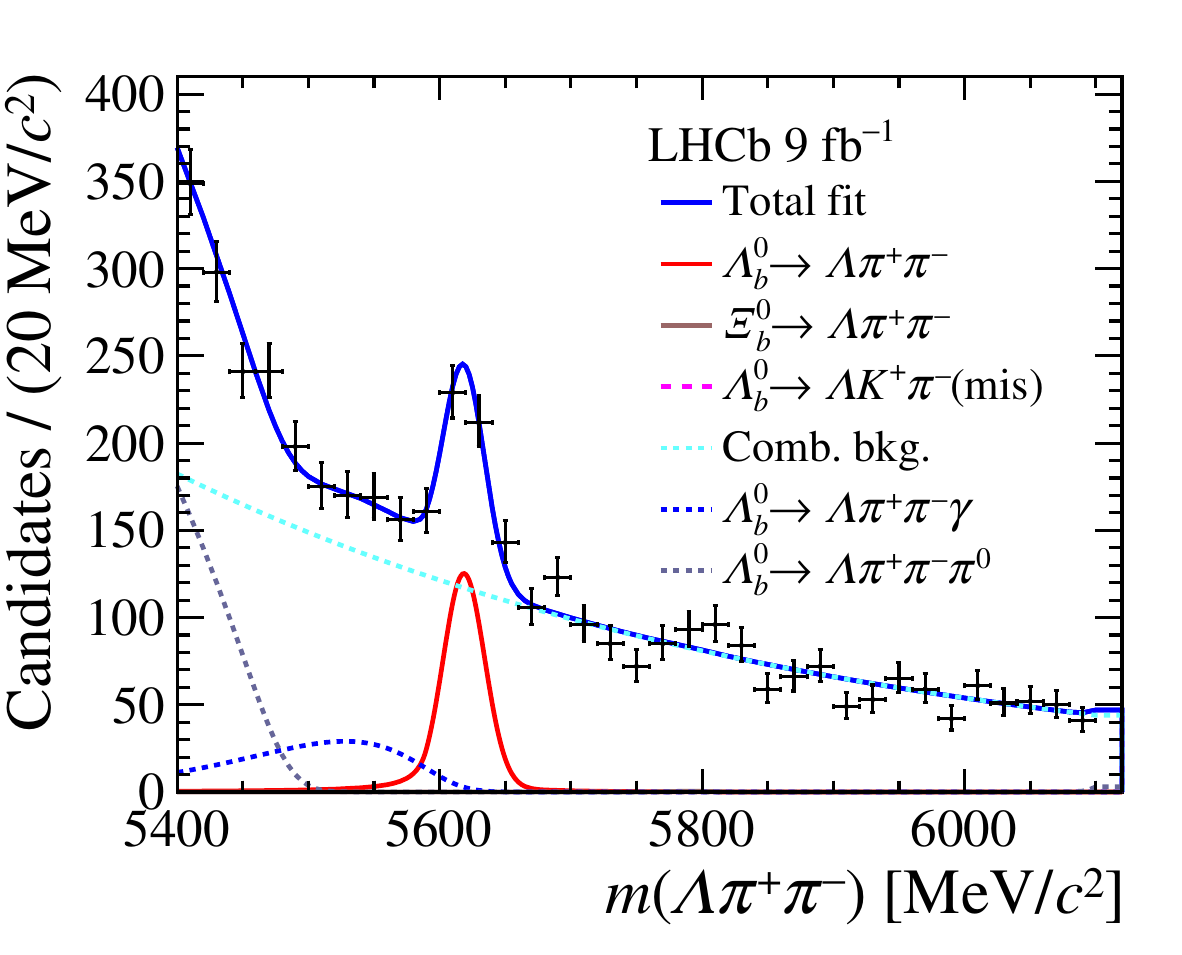}\put(-145,128){(c)}
    \includegraphics[width=0.44\columnwidth]{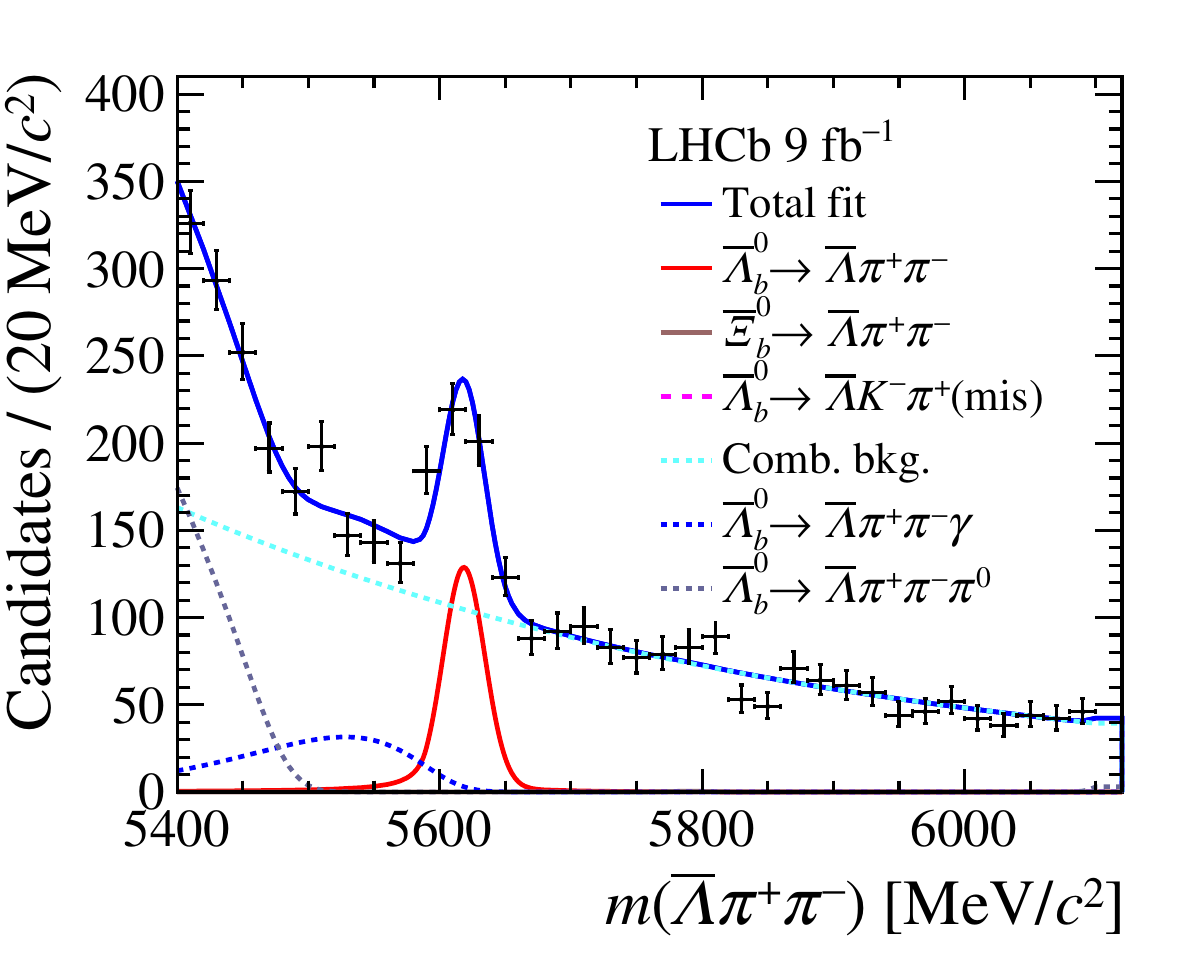}\put(-145,128){(d)}\\
    \includegraphics[width=0.44\columnwidth]{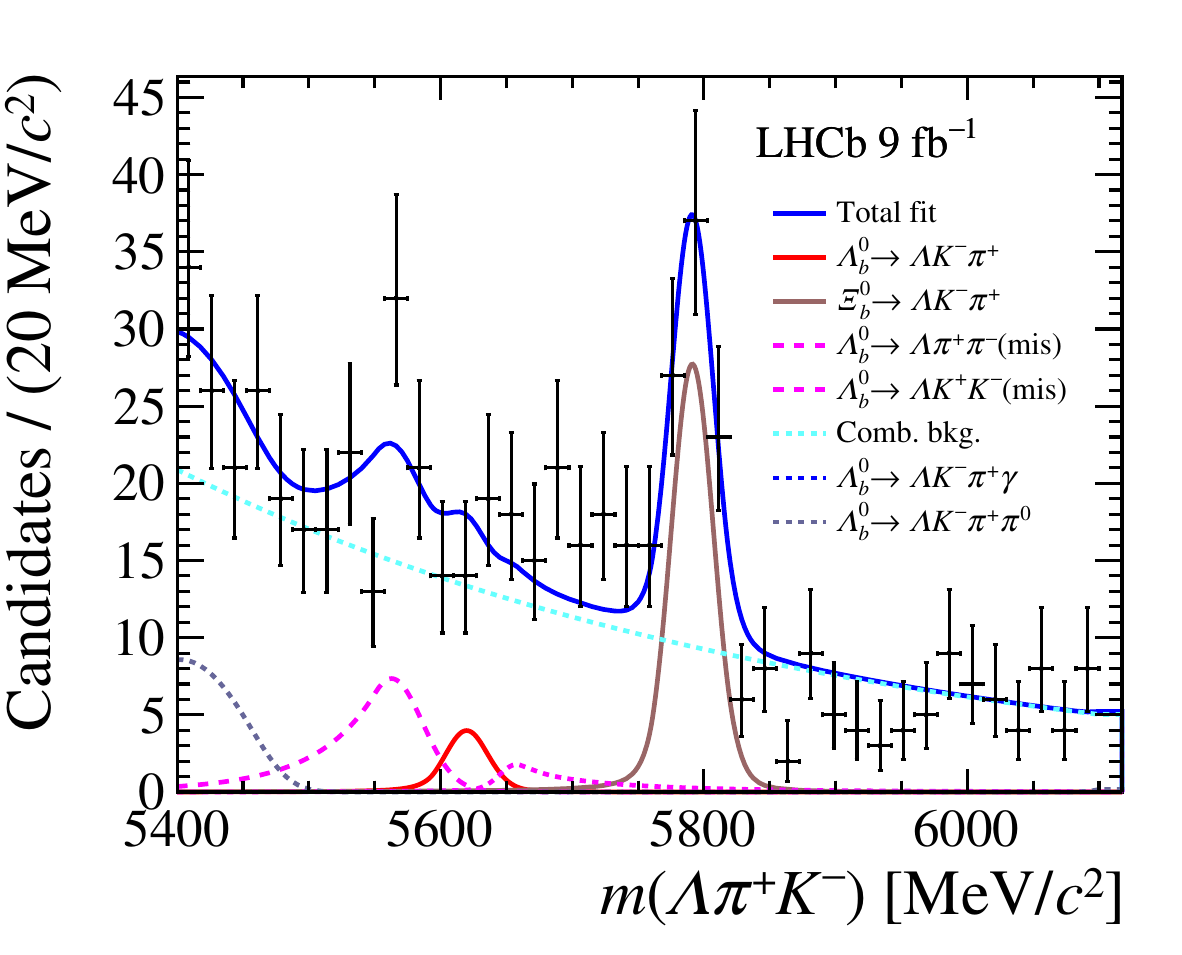}\put(-145,128){(e)}
    \includegraphics[width=0.44\columnwidth]{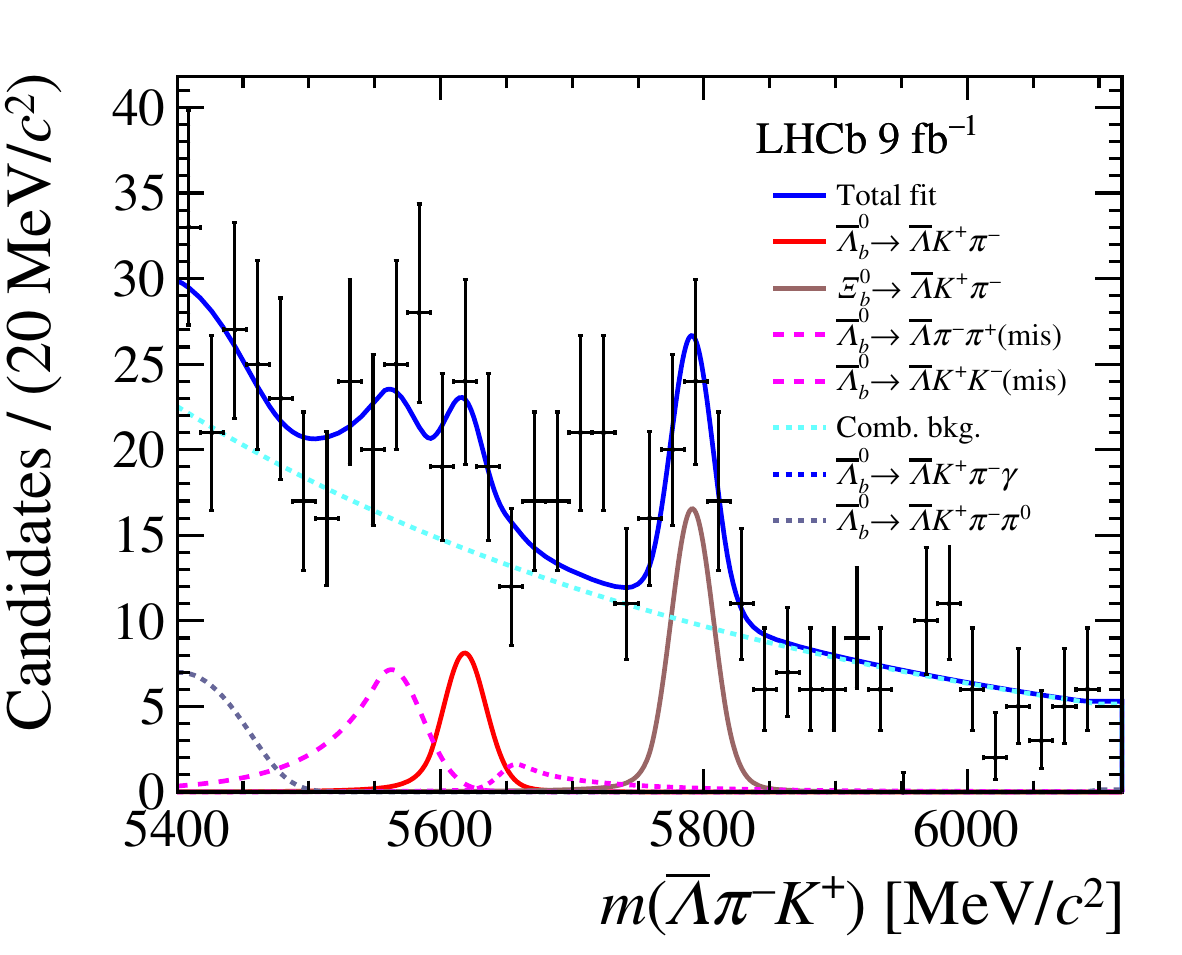}\put(-145,128){(f)}

  \end{center}
  \caption{Distributions of $m(\Lz h^+h^-)$ for (a) $\decay{\Lb}{\Lz K^+\pi^-}$, (b) $\Lbbar\to\Lbar \Km\pip$, 
    \mbox{(c) $\decay{\Lb}{\Lz \pi^+\pi^-}$}, (d) $\Lbbar\to\Lbar \pi^+\pi^-$, (e) $\decay{\Xibz}{\Lz K^-\pi^+}$, and (f) $\Xibbarz\to\Lbar\Kp\pim$
    decays in data, together with the fit results. The selection criteria are optimized with the $N_{\rm{S}}/\sqrt{N_{\rm{S}} + N_{\rm{B}}}$ FoM for \Lb distributions and $N_{\rm{S}}/(\sqrt{N_{\rm{B}}} + 2.5)$ FoM for \Xibz distributions.  }
  \label{fig:ACP}
\end{figure}

Figure~\ref{fig:mass_ref} shows the mass spectrum of the control channel $\Lb\to\Lc(\to\Lz\pip)\pim$ and $\Lbbar\to\Lcbar(\to\Lbar \pim)\pip$.
Figure~\ref{fig:ACP} shows the mass spectra of the \LzKP, \LzPP and \XiPK decays, separately for baryon and antibaryon samples.
Due to the relatively smaller number of $\Xibz$ signal yields, when determining its selection criteria the $N_{\rm{S}}/(\sqrt{N_{\rm{B}}} + 2.5)$ FoM method is applied, as shown in Fig.~\ref{fig:ACP} (e)(f), whereas when studying the $\Lb$ modes, the $N_{\rm{S}}/\sqrt{N_{\rm{S}} + N_{\rm{B}}}$ FoM method is applied, as shown in Fig.~\ref{fig:mass spectrum} (a)(b)(c)(d).

\begin{comment}

\end{comment} 

\end{document}